\documentclass{aa}  

\usepackage{placeins}

\usepackage{graphicx}
\usepackage{amssymb}
\usepackage{lipsum}
\usepackage{subcaption}
\usepackage{csquotes}

\usepackage{txfonts}

\begin{document}

   \title{Challenging the archetypal intermittent radio galaxy J0111+3906. New insights from optical GTC/OSIRIS and radio VLA and LOFAR observations}

   \author{M. Orienti
          \inst{1}\
          \and
          J.A. Acosta-Pulido\inst{2,3}
          \and
         E. De Rubeis\inst{4,1}
\and
         C. Stanghellini\inst{1}
         \and
         M. Bondi\inst{1}
         \and
         F. D'Ammando\inst{1}
             \and
        L. Coccato\inst{5}
             \and
        I. Alemán-Mart\'in\inst{3}
}
   \institute{INAF - Istituto di Radioastronomia, Via P. Gobetti 101, I-40129 Bologna, Italy\\
              \email{orienti@ira.inaf.it}
        \and
Instituto de Astrof\'isica de Canarias, Calle Vía Láctea, s/n, E-38205, La Laguna, Tenerife, Spain
\and
Departamento de Astrof\'isica, Universidad de La Laguna, E-38206, La Laguna, Tenerife, Spain
        \and
        Hamburger Sternwarte, Universit\"at Hamburg, Gojenbergsweg 112, 21029 Hamburg, Germany       
        \and
        European Southern Observatory, Karl-Schwarzschild-Strasse 2, D-85748, Garching, Germany             }

   \date{Received <date> / Accepted <date>}

 
  \abstract
      {
We present results on new optical Gran Telescopio CANARIAS OSIRIS+ spectroscopy and radio Very Large Array (VLA) observations at 360 MHz of the optical object behind the radio emission 20 arcseconds east of the archetypal restarted compact symmetric object (CSO) J0111+3906. The optical counterpart is a post-starburst galaxy at redshift $z$ = 1.042, ruling out any physical relation between this object and the CSO radio galaxy J0111+3906 at redshift $z$ = 0.668. The radio emission is therefore not the remnant of a past activity of the CSO J0111+3906 that took place around 10$^7$ years ago. When observed with the high angular resolution of VLA and the International LOw Frequency ARray telescope
the East radio source is resolved in a compact component with spectral index 0.7, centred on the optical galaxy, surrounded by diffuse emission with a slightly steeper spectrum. The upper limit to the radiative age of the East radio source is a few million years. If the trigger of the radio emission and the recent burst of star formation observed in the East radio source are causally connected, there must be a long time gap between the two phenomena. The multi-frequency analysis of the CSO radio galaxy J0111+3906 pointed out excess flux density at 150 MHz, which may indicate the presence of fossil plasma confined within the host galaxy. In this scenario, the past activity of the CSO J0111+3906 switched off not long after its trigger, suggesting a duty cycle of about 10$^4$ yr.}

\titlerunning{Optical and radio observations of  J0111+3906}
\authorrunning{M. Orienti et al.}
   \maketitle
   
\nolinenumbers
 
%

\section{Introduction}

Compact symmetric objects (CSOs) are extragalactic radio sources of linear size (LS) $<$ 1 kpc and a two-sided structure that is reminiscent of Fanaroff-Riley radio galaxies on smaller scales. In the evolutionary scenario of powerful jetted active galactic
nuclei (AGNs), the linear size of a radio source is related to the age of the radio emission. In this context, CSOs likely represent radio sources in an early evolutionary stage \citep[e.g.][]{fanti95,readhead96,odea21}, although alternative scenarios have been proposed \citep[e.g. jet frustration and/or redirection,][]{cstan25}. 
Estimates of
kinematic and radiative ages of $10^{2-4}$ yr \citep[e.g.][]{murgia03,polatidis03,an12a} support
an evolutionary path in which CSOs are the progenitors of classical radio sources with typical ages of $10^{7-8}$ yr \citep[e.g.][]{orru10}.

\indent The excess of young radio sources in flux-limited samples
may indicate the existence of short-lived objects unable to evolve into
large radio sources. The
overabundance of CSOs with ages less than 500 yr is consistent with
this scenario \citep{gugliucci05,an12b,kiehlmann24}. 
The motivation
behind short-lived objects is still highly debated: instabilities in
the accretion disks, tidal disruption events, and extraction of the spin
energy of the supermassive black hole are all viable mechanisms \citep{czerny09,readhead24}.

\indent After the relativistic jet switched off, it is possible that another
epoch of radio emission began. Double-double radio sources with two (or more) pairs of radio lobes are clear examples of galaxies with recurrent jet activity \citep[e.g.][]{lara99,schoenmakers00,brocksopp07,kuzmicz17}. Steep-spectrum diffuse radio emission associated with CSOs has been found on milliarcsecond scales \citep[tens of parsecs;][]{mo08,mo23} and on
arcsecond scales \citep[i.e. from tens to
hundreds of kiloparsecs;][]{cstan98,cstan05} pointing out duty cycles
of the order of $10^{3-4}$ and $10^{7-8}$ yr, respectively. \\

\begin{figure}
\includegraphics[width=0.97\columnwidth]{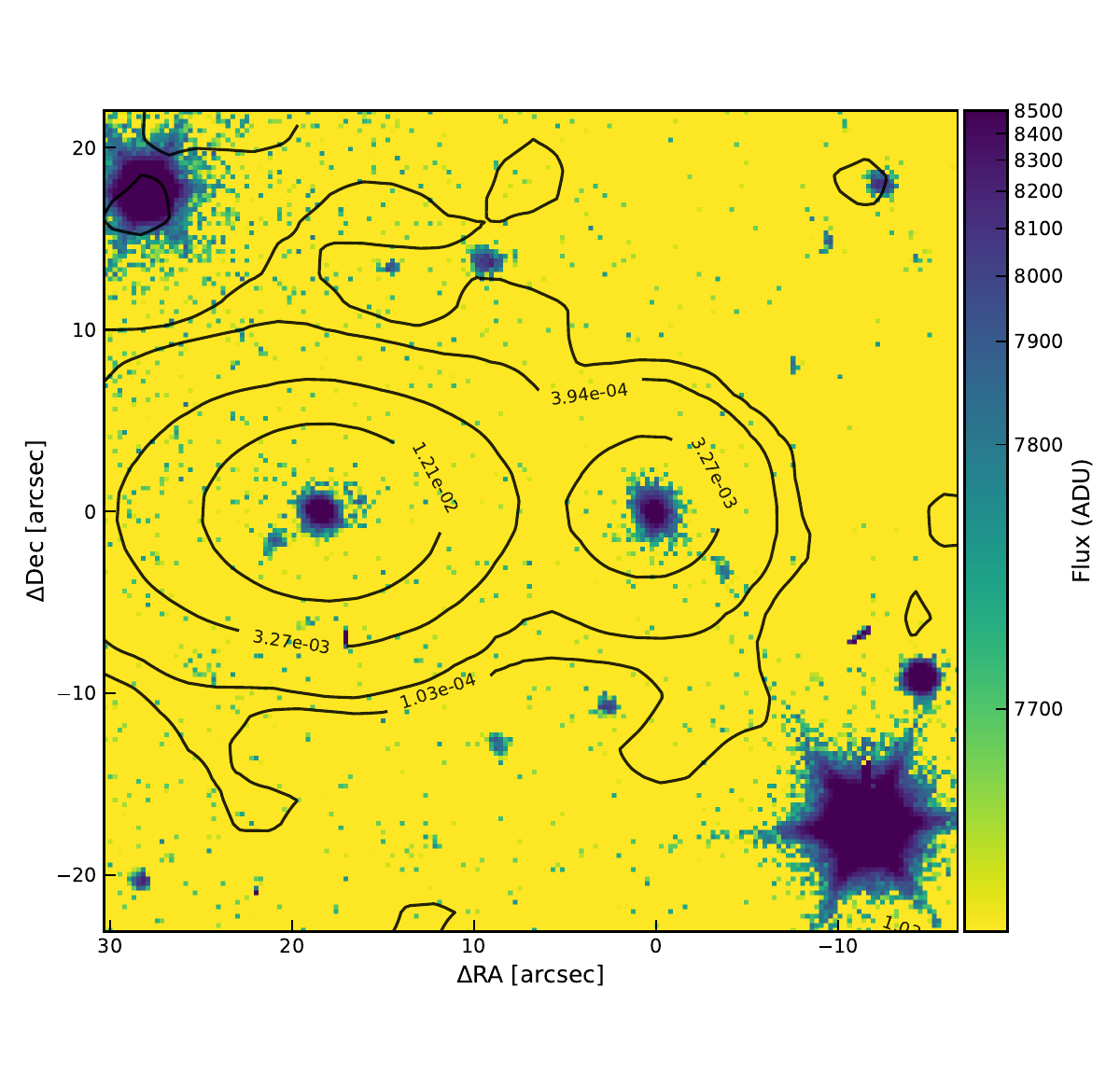}
\caption{LOFAR contours at 150 MHz of J0111+3906 overlaid the optical SDSSr image taken with GTC/OSIRIS.  The orientation is north up and east left. The optical counterpart of the East radio source appears rather compact and shows a plume at $\sim 3"$\,SE.}  
\label{lofar-optical}
\end{figure}

\indent The radio galaxy J0111+3906 is the archetype of restarted CSOs \citep{baum90}. This radio source, hosted by a galaxy at $z=0.67$ \citep{lawrence96}, shows a two-sided radio structure
with LS$\sim$40 pc and an estimated kinematic age of about 400 yr \citep{owsianik98}. Diffuse emission was discovered about $20''$ east of the CSO, showing a misalignment of about 30$^{\circ}$ with respect to the parsec-scale radio structure. No strong claim on the spectral index of the kiloparsec-scale component was made, owing to the large uncertainty on the flux density, mainly at low frequencies, where the extended component and the CSO could not be well resolved by available data \citep{baum90}. \\
Despite the lack of spectral information and the misalignment between parsec and kiloparsec scales, the arcsecond-scale diffuse emission has been considered a relic of past nuclear activity of the radio galaxy J0111+3906. However, new observational evidence raises doubts about this hitherto accepted interpretation. Deep optical images point out the presence of a galaxy 
at the position of diffuse emission (RA: 01:11:38.8, Dec=+39:06:27.5,
J2000, Fig. \ref{lofar-optical}), suggesting a scenario in which the extended radio component (hereafter East radio source) may be unrelated to the CSO 
\citep{cstan25}.\\

In this paper, we present results on new Jansky Very Large Array (VLA) radio and Gran Telescopio CANARIAS (GTC) optical observations of both the CSO J0111+3906 and the East radio source. Archival International LOw Frequency ARray (LOFAR) Telescope ~\citep[ILT,][]{vanhaarlem2013} data complement our observations with the aim of investigating the physical properties of the East radio source. This will allow us to confirm or disprove the intermittency of the radio emission in the CSO radio galaxy J0111+3906. \\
 This paper is organized as follows: Sect. \ref{sec-obs} describes observations and data analysis, while the results are presented in Sect. \ref{sec-results} and discussed in Sect. \ref{discussion}. In Sect. \ref{summary} we draw our conclusions.\\

Throughout the paper, we assume the following cosmology:
$H_{0} =70\; {\rm km\,s^{-1}\, Mpc^{-1}}$, $\Omega_{\rm M} = 0.27,$ and $\Omega_{\rm \Lambda} = 0.73$ in a flat Universe. 
The spectral index, $\alpha,$ is defined as S($\nu$) $\propto \nu^{- \alpha}$. 

\section{Observations and data analysis}
\label{sec-obs}

\subsection{Optical data - GTC/OSIRIS observations}

J0111+3906 was observed on 30 September, 2024, using the OSIRIS+ instrument \citep{Cepa03} mounted on the GTC Cassegrain focus. Observations were taken under photometric conditions and good and stable seeing (full width at half maximum (FWHM)~$\simeq 0\farcs 8$). Spectra were taken using the R1000R grism, which covers the spectral range 5100-10000~$\AA$. The spectral range beyond 9000~$\AA$ is strongly affected by sky OH emission lines and telluric absorption. Six exposures of 1200 s were taken, giving a total exposure time of 2 hours. The 1.2 arcsec width slit was selected, giving a resolution of $\sim 390$ km/s. The slit was oriented at ${\rm PA}=-89.5^\circ$ in order to have both targets within it. The parallactic angle varies from PA = 150$^\circ$ at the beginning of the observations to $\rm{PA}=99^\circ$ at the end, with the airmass changing from 1.021 to 1.152. Differential atmospheric refraction (DAR) flux loss effects remain very small during the observations given the low airmass and the slit width as compared with the seeing value.\footnote{DAR was computed in the spectral range from 5000 to 9000~$\AA$ for the minimum and maximum values of the airmass, resulting in $0\farcs 179$ and $0\farcs 497$ at an airmass of 1.021 and 1.152, respectively. Flux losses due to DAR occur only perpendicularly to the slit, resulting in offsets of $0\farcs 155$ and $0\farcs 079$ at the beginning and end of the observations. The corresponding slit losses range from 3\% to 1\%. }
 Data were reduced using the PypeIt v1.15.1 pipeline \citep{Prochaska20a,Prochaska20b}. The data reduction process includes the following steps: bias subtraction, flat-field, sky subtraction, and optimal extraction. Finally, individual spectra of each target were coadded. The spectrophotometric standard star Ross 640 was used to flux calibrate the spectra. 

\begin{figure}
\includegraphics[width=1.25\columnwidth]{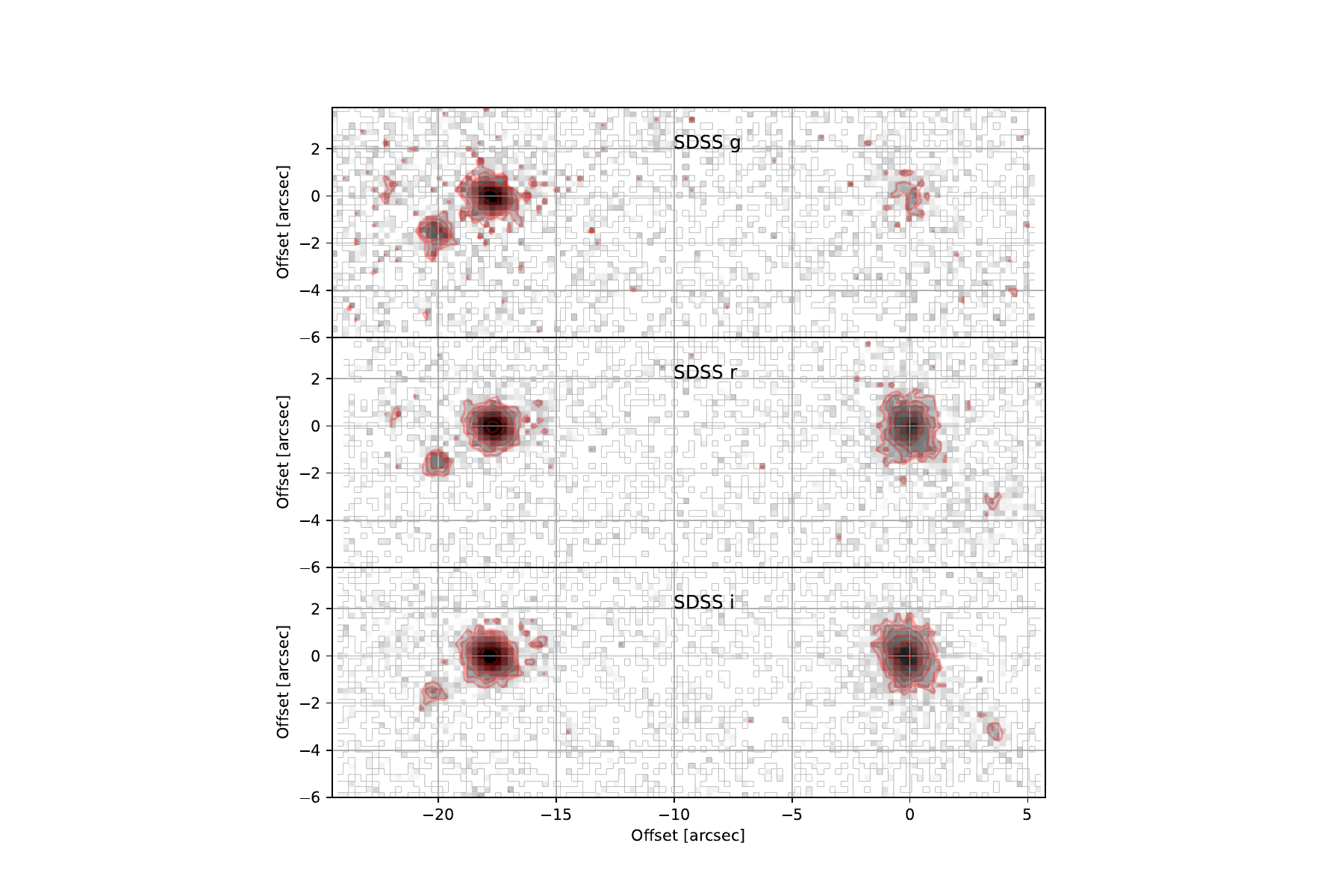}
\caption{Image zoomed around the CSO. The  field is shown in the filters SDSSg, SDSSr, and SDSSi. The orientation is north up and east  left.}  
\label{fi:osiris_zima}
\end{figure}

In addition, images were taken through the SDSS filters g, r, and i, with  exposure times of 200\,s, 150\,s, and 100\,s,   respectively. Data were reduced using modules of astropy.ccdred. The flux calibration was performed using several reference stars with available PanSTARRS photometry and no signs of variability as indicated by the $\sigma$ of the {\it MeanPSF} in all SDSS filters.  A zoom of the images close to the CSO  is shown in Fig \ref{fi:osiris_zima}. Note that the CSO is barely detected in the SDSSg filter, whereas it is clearly detected at SDSSr and SDSSi. The East radio source appears rather compact. There is a small blob at the south-east of the East radio source, which has similar colours and could be a close-by galaxy. The photometry results are listed in Table \ref{ta:opt_phot}.

\begin{table}
\caption{Optical photometry.}
\begin{center}
\begin{tabular}{lccclc}
\hline
Filter & CSO & East source  \\
       & mag & mag        \\
      \hline
SDSSg  & 22.83 $\pm$ 0.25   & 21.65 $\pm$ 0.23 \\
SDSSr  & 20.95 $\pm$ 0.14   & 20.80 $\pm$ 0.14 \\
SDSSi  & 20.10 $\pm$ 0.11   & 20.05 $\pm$ 0.11 \\
\hline
\end{tabular}
\end{center}
\label{ta:opt_phot}
\tablefoot{Optical photometry of the CSO J0111+3906 and the East radio source.}
\end{table}

\subsection{Radio data}

\begin{table*}
\caption{Image parameters.}
\begin{center}
\begin{tabular}{lccclc}
\hline
Array & $\nu$ & rms & S$_{\rm p}$ & \;\;\;\;\;\;\;\;\;\;\; Beam & Weighting \\
      & GHz   & mJy/beam & mJy/beam & arcsec $\times$ arcsec \;\; deg & \\
      \hline
LOFAR & 0.15 & 0.085 & 35.0 & 6.00 $\times$ 6.00 \;\;\;\;\;\;\;\; 90 & - \\
ILT & 0.15 & 0.065 & 14.0 & 0.38 $\times$ 0.34 \;\;\;\;\;\;\;\;\;\; 1 & - \\
VLA   & 0.368 & 0.39 & 24.7 & 4.47 $\times$ 3.19 \;\;\;\;\;\; $-$81 & U \\
VLA   & 0.368 & 0.24 & 24.7 & 5.23 $\times$ 3.91 \;\;\;\;\;\; $-$87 & B \\
VLA   & 0.352 & 0.33 & 25.7 & 8.48 $\times$ 6.26 \;\;\;\;\;\; $-$89 & N \\
VLA   & 1.5   & 0.13 & 504.6 & 3.31 $\times$ 2.84 \;\;\;\;\;\;\;\; 45 & U \\
VLA   & 1.5   & 0.065 & 504.6 & 4.05 $\times$ 3.65 \;\;\;\;\;\;\;\; 41 & B \\
VLA   & 1.5   & 0.065 & 504.6 & 5.05 $\times$ 4.63 \;\;\;\;\;\;\;\; 31 & N \\
VLA   & 4.8   & 0.030 & 1297.0 & 0.34 $\times$ 0.32  \;\;\;\;\;\; $-$79 & U \\
VLA   & 4.8   & 0.035 & 1299.0 & 3.00 $\times$ 2.80 \;\;\;\;\;\; $-$74 & N \\
\hline
\end{tabular}
\end{center}
\label{radio-oss-tab}
\tablefoot{Column 1: Array. Column 2: Frequency; Column 3: 1$\sigma$ rms noise level. Column 4: Peak flux density. Column 5: Clean beam size and position angle of the major axis. Column 6: Weighting used to create the image: B: Briggs (robust=0), N = natural, U: uniform.}
\end{table*}

\subsubsection{VLA observations}

VLA observations of J0111+3906 in the P band (central frequency 368 MHz) were carried out with the array in A-configuration for a total observing time of 6 h (project code VLA/24B-144). The observations were carried out during nighttime and split into three sessions (10 December, 2024, 29 December, 2024, 4 January, 2025) to avoid ionospheric phase fluctuations during sunset and sunrise. The observations were performed with an 8-bit continuum setup and a 256-MHz band width. About 40\% of the total band width was affected by radio frequency interference (RFI). The source 3C\,48 was used as primary, band pass, and secondary calibrator. During each observing block, the target source was observed for a total of about 75 min, divided into 3 scans of 25 min each, interleaved by 2-min scans on 3C\,48. 

The calibration of each dataset was performed using Common Astronomical Software Applications \citep[CASA,][]{mcmullin07} version 6.7 following the standard procedure for VLA observations at low frequencies. 
Data were inspected and Hanning smoothed to prevent Gibbs ringing caused by severe RFI. Automated flagging was performed on bad data, followed by an initial delay and band pass calibration of 3C\,48. Afterwards, we performed a second flagging run. Then we did the entire calibration on the flagged data. After checking for antenna position corrections and ionospheric total
electron content corrections, we set the flux density model for 3C\,48 using the \citet{scaife12} scale. Then we performed delay, band pass, and gain calibration on 3C\,48. As a last step, we transferred the calibrations to the target field. A final flagging run was performed on the calibrated dataset of the target field. 
The error in amplitude calibration, $\sigma_{\rm cal}$, was estimated by checking the scatter of the amplitude gain factors and turned out to be about 5\%, in agreement with what was found by \citet{pb17}. The calibrated datasets were concatenated to improve the signal-to-noise ratio. 

We produced several sets of images using the CASA task {\tt tclean} with multi-term multifrequency synthesis deconvolution (nterms=2), and assuming Briggs (robust=0), uniform, and natural weightings (Fig. \ref{vla-fig}). The image parameters are reported in Table \ref{radio-oss-tab}.\\

We complement our observations in the P band with archival data in the L and C bands. The log of VLA observations is reported in Table \ref{vla-log}. 

L band observations (central frequency 1.5 GHz) were performed on 15 January, 2023 when the VLA was in A-configuration (project code VLA/23A-006). J0111+3906 was the phase calibrator of the project and was observed for a total time of about 15 min. We retrieved the calibrated dataset from the archive and inspected the data to check whether additional flagging and/or calibration steps were needed. The uncertainty in gain calibration was conservatively assumed to be about 5\%. The final images produced with different weightings are shown in Fig. \ref{vla-fig}.

Several datasets in C band acquired with the historical VLA and covering all VLA configurations were retrieved from the National Radio Astronomy Observatory (NRAO) archive. Calibration and hybrid mapping were performed using standard procedures described in the Astronomical Image Processing System (AIPS) cookbook. The longest observations (about 9 hours) were carried out with the array in the A-configuration and were imaged separately to disentangle the contribution of the compact region at the centre of the East radio source. A slight offset in flux density, due to uncertainties in the a priori calibration, and conservatively estimated at 5\%, was corrected by normalizing the antenna gains of all datasets to those of the A-configuration data. The images are shown in Fig. \ref{vla-cband}.\\

\begin{figure*}
\includegraphics[width=\columnwidth]{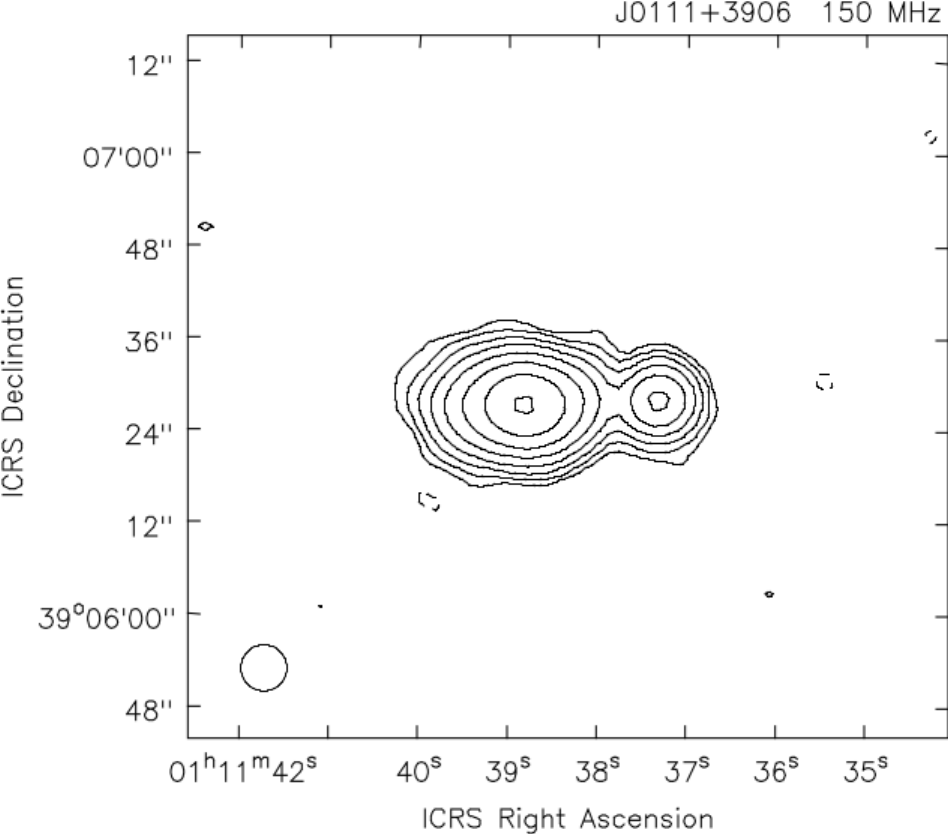}
\includegraphics[width=\columnwidth]{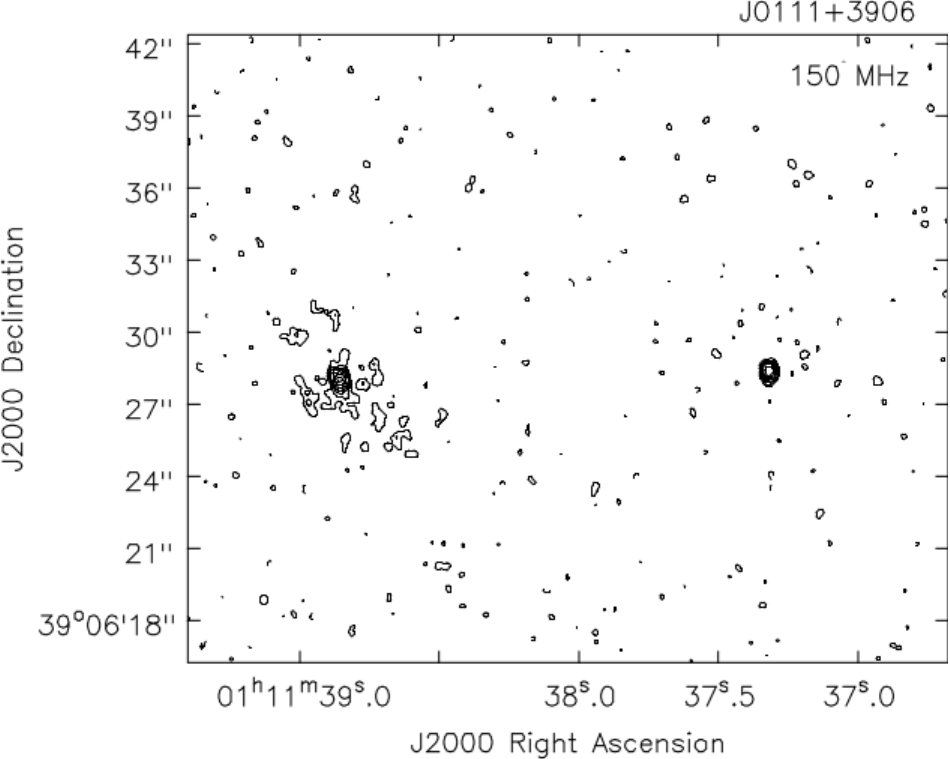}
\caption{Image at 150 MHz of the CSO J0111+3906 and the East radio source obtained with LOFAR ({\it left}) and with ILT ({\it right}). Image parameters are reported in Table \ref{radio-oss-tab}. The first contour is three times the rms. Contours increase by a factor of 2. The restoring beam is plotted in the bottom left-hand corner of each image.}   \label{lofar-fig} 
\end{figure*}

\subsubsection{ILT data}
\label{sec-lofar-ilt}
J0111+3906 was observed with LOFAR in 2019 (project: LC9\_030, P.I. Shimwell) as a pointing of the LOFAR Two-Metre Sky Survey~\citep[LoTSS,][]{shimwell2022}. The observation was made in high-band antenna (HBA, $120-168$~MHz) for 8 hours, using 13 international stations (IS), which allow sub-arcsecond resolution. LOFAR Initial Calibration Pipeline~\citep[LINC,][]{vanweeren2016,williams2016,degasperin2019} was used on the primary calibrator (3C\,295) to correct for the polarization alignment, clock delays between stations and the bandpass, both for Dutch and IS. The LINC was also used to correct for direction-independent effects (DIE) for the target field on Dutch stations, using the TIFR GMRT Sky Survey~\citep[TGSS,][]{intema2017} as the initial sky model. The IS calibration was performed using the LOFAR-VLBI pipeline~\citep[][van der Wild \textit{in preparation}]{morabito2022}. As delay-calibrator, a source required to correct for direction-independent dispersive delays, we used source B3\,0107+387~\citep{ficarra1985} from the Long Baseline Calibrator Survey~\citep[LBCS,][]{jackson2016,jackson2022}. After applying DIE solutions to the IS, the data are concatenated into measurement sets of 2~MHz bandwidth each, and phase-shifted towards the direction of the delay-calibrator. Delay calibration is performed through facetselfcal~\citep{vanweeren2021}, which enables self-calibration routines that combine Default PreProcessing Pipeline (DP3) for calibration and WSClean~\citep{offringa2014,offringa2017} for imaging. To constrain the astrometric accuracy of the source, we used a snapshot from the VLA Sky Survey~\citep[VLASS,][]{lacy2020} as an initial sky model. To reduce interference from an unrelated nearby radio source, LOFAR’s core stations were digitally phased into a single superstation, data were averaged to 16~s time integrations and 195~kHz frequency channels, and only baselines longer than $40~\rm{k \lambda}$ were used. After self-calibration on the delay calibrator, the data were phase-shifted towards the direction of J0111+3906 and solutions from the delay calibration were applied. Due to its faintness, no self-calibration was performed on the target source. The final image, produced with WSClean, is shown in the right panel of Fig.~\ref{lofar-fig}.

\section{Results}
\label{sec-results}

\subsection{The optical emission}

\begin{figure}
\includegraphics[width=0.99\columnwidth]{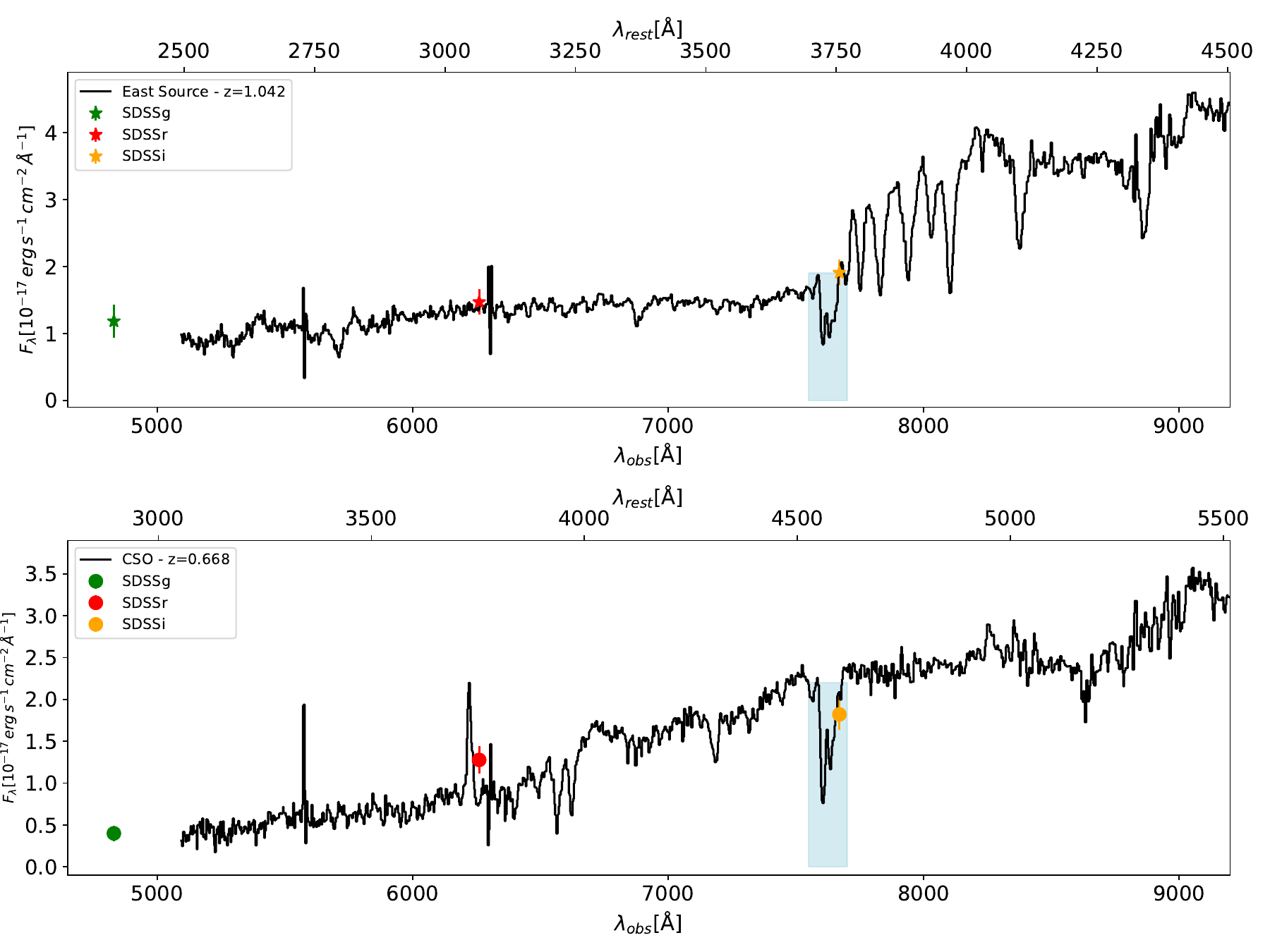}
\caption{GTC/OSIRIS spectra extracted from the positions of the East radio source (top panel) and the CSO (bottom panel). Both spectra were corrected by Galactic dust reddening using the model introduced by \citet{Gordon23} with $A_V=0.129$.
The coloured symbols, green, red and orange, correspond to the photometry at the filters SDSSg, SDSSr, and SDSSi, respectively. Photometry was corrected by Galactic dust reddening using values retrieved from the NASA/IPAC Extragalactic Database. Shaded regions correspond to the telluric absorption wavelength range. The top x-axis indicates wavelength in the rest frame.}  
\label{fi:osiris-2spectra}
\end{figure}

The spectra extracted at the positions of the CSO and the East radio source are shown in Fig \ref{fi:osiris-2spectra}. The photometric measurements obtained from the SDSS images (Fig. \ref{ta:opt_phot}) are overlaid in this figure. These images were  taken with OSIRIS+ quasi-simultaneously with the spectra, and they were used to cross--check the photometric calibration of the spectra. The \ion{Ca}{II} K \& H doublet was recognized in the two spectra and allows us to measure the redshift in a secure way. Thus, the redshift of the CSO is confirmed to be $z=0.6685 \pm 0.0003$\footnote{At $z = 0.668$, 1 arcsecond corresponds to 7.106 kpc and the luminosity distance is D$_{\rm L} = 4079.7$ Mpc.} and for the first time the redshift of the East radio source could be measured, resulting in a value $z=1.0421 \pm 0.0002$\footnote{At $z = 1.042$, 1 arcsecond corresponds to 8.236 kpc and the luminosity distance is D$_{\rm L} = 7084.6$ Mpc.}. A more detailed description is given in the following subsections. 

\subsubsection{The CSO spectrum}
\label{ssc:cso_spect}

\begin{figure*}
\includegraphics[width=1.95\columnwidth]{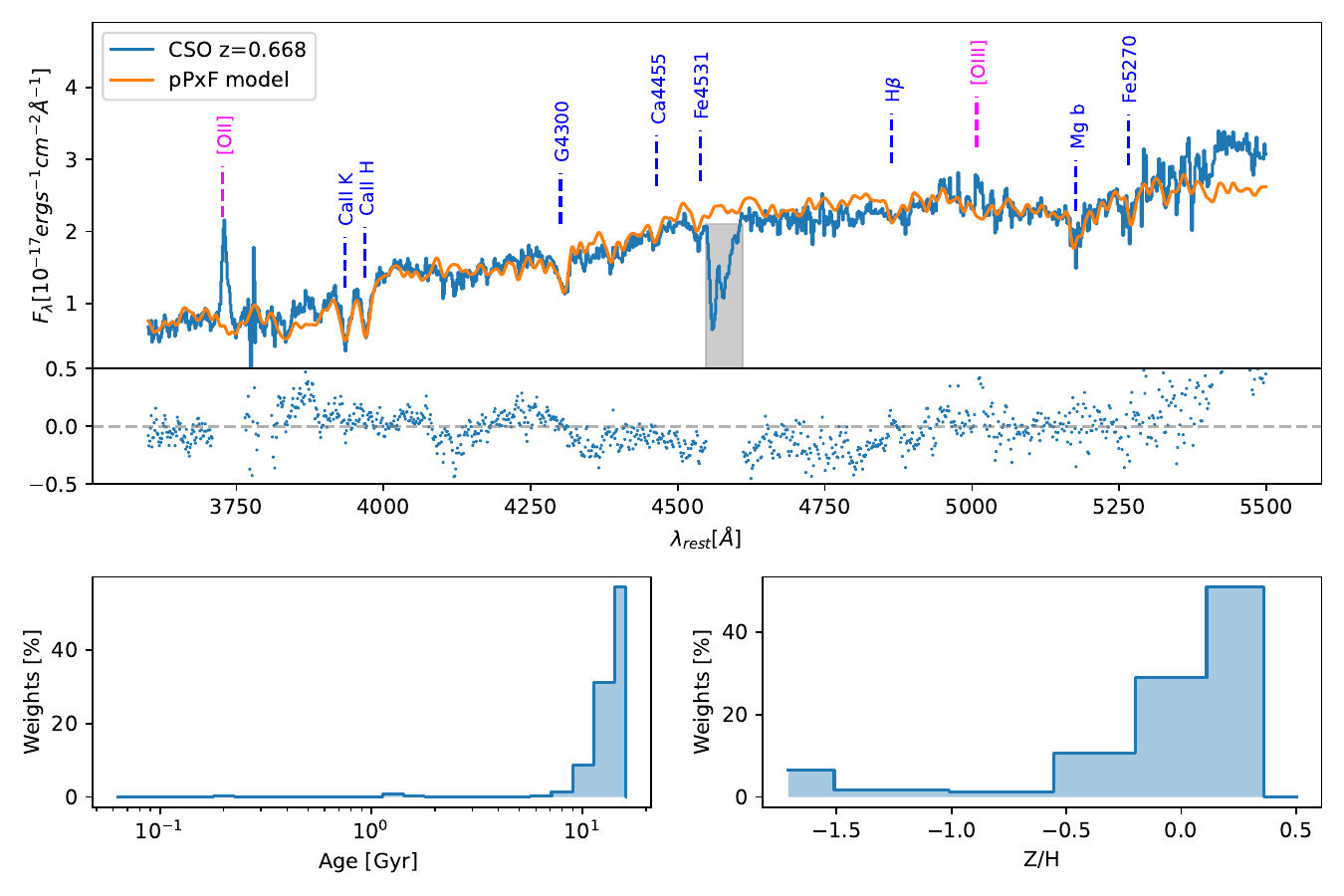}
\caption{pPXF model of the CSO optical GTC/OSIRIS spectrum. The top panels show the observed spectrum extracted at the CSO position (shifted to the wavelength rest frame) together with the best fit model obtained using pPXF. The shaded region indicates the telluric absorption. Residuals are also presented in the attached sub-panel. The bottom panels represent the luminosity-weights of a single stellar population (SSP) for a given age (bottom left panel) and for a given metallicity (bottom right panel). Note the good matching between the composite stellar templates and the observed spectrum. The emission line [\ion{O}{II}]$\lambda 3727$ \AA\ is clearly visible and the [\ion{O}{III}]$\lambda 5007$ \AA\ can be glimpsed. }  
\label{fi:ppxf_cso}
\end{figure*}

The GTC/OSIRIS+ spectrum extracted at the CSO location shows several conspicuous absorption features that are identified as \ion{Ca}{II} K \& H, the G-band [4300$\AA$], and the Mg-b features (see Figs. \ref{fi:osiris-2spectra} and \ref{fi:ppxf_cso}). The H Balmer lines appear very shallow, indicating an old stellar population. In contrast, there is a prominent emission line in the blue end of the spectrum, which corresponds to [\ion{O}{II}]$\lambda 3727 \AA$. However, the commonly observed line [\ion{O}{III}]$\lambda 5007 \AA$ appears rather faint with a low S/N ratio. Both lines were measured and the results are presented below. 
 
The spectrum was modelled using the pPXF software \citep{Cappellari2023}. We used
stellar templates from the eMILES library \citep{Vazdekis2016}, with the Padova isochrones \citep{Girardi2000} and the Salpeter initial mass function (IMF). These templates span a range of 25 ages from 63 Myr to 15.8 Gyr, and 6 [Z/H] values from $-$1.71 to 0.22. 
The spectrum was corrected by Galactic dust reddening using the model introduced by \citet{Gordon23} with $A_V=0.129$. The emission line and the telluric absorption features were masked during the fit. The results after applying the pPXF model are represented in Fig. \ref{fi:ppxf_cso}. The best fit model reproduces reasonably well the observed spectrum. 
This model corresponds to an old stellar population ($\simeq 13 \, {\rm Gyr}$) with solar and supersolar metallicity. We notice that the observed optical continuum can be explained uniquely by the stellar population, and any extra contribution, for example from the active nucleus, appears negligible. Therefore, it can be concluded that the optical emission from the CSO is entirely dominated by the host galaxy emission.

\subsubsection{The CSO emission line luminosity}
\label{sec:emission_line_luminosity}
The emission lines [\ion{O}{II}]$\lambda 3727$ and [\ion{O}{III}]$\lambda 5007$ were measured from the spectrum after subtracting the stellar contribution. 
We used the module {\it line\_flux} from the package {\it specutils.analysis} to measure the line flux. 
The results are $F[\ion{O}{II}] = (3.27 \pm 0.17) \times 10^{-16} {\rm erg\,s^{-1}\,cm^{-2}}$  and $F[\ion{O}{III}] = (1.23 \pm 0.14) \times 10^{-16} {\rm erg\,s^{-1}\,cm^{-2}}$. The ratio [\ion{O}{II}]/[\ion{O}{III}] indicates low ionization, which classifies this radio source as a low ionization nuclear emission-line region (LINER). The corresponding luminosity is ${\rm L[\ion{O}{II}]}= (6.70 \pm 0.35) \times 10^{41} {\rm erg\,s^{-1}}$ and ${\rm L[\ion{O}{III}]}= (2.53 \pm 0.29) \times 10^{41} {\rm erg\,s^{-1}}$. 
If the [OII] emission is produced by ionization by hot stars, the star formation rate (SFR) can be estimated by following the expression: ${\it SFR}[M_\odot yr^{-1}]= 1.4\times 10^{-41} \times {\rm L[\ion{O}{II}](\mathrm{erg\,s^{-1}})}$ \citep{Kennicut98}. 
This expression yields ${\it SFR} = 9.4 M_\odot \; {\rm yr^{-1}}$,  which is a moderate star-formation rate. However, this value seems inconsistent with the stellar population history from the pPXF modelling, indicating the AGN as the origin of [\ion{O}{II}] and [\ion{O}{III}] emission lines. Assuming the AGN as the ionization source, the bolometric luminosity (${\rm L_{iso}}$) can be estimated using the relationships provided by \cite{Pennell17}, and results in values ${\rm L_{iso}}= (0.8,2.8) \times 10^{45} {\rm erg\,s^{-1}}$ using traditional or improved corrections, respectively. 

\subsubsection{The optical spectrum of the East radio source}

\begin{figure*}
\includegraphics[width=1.95\columnwidth]{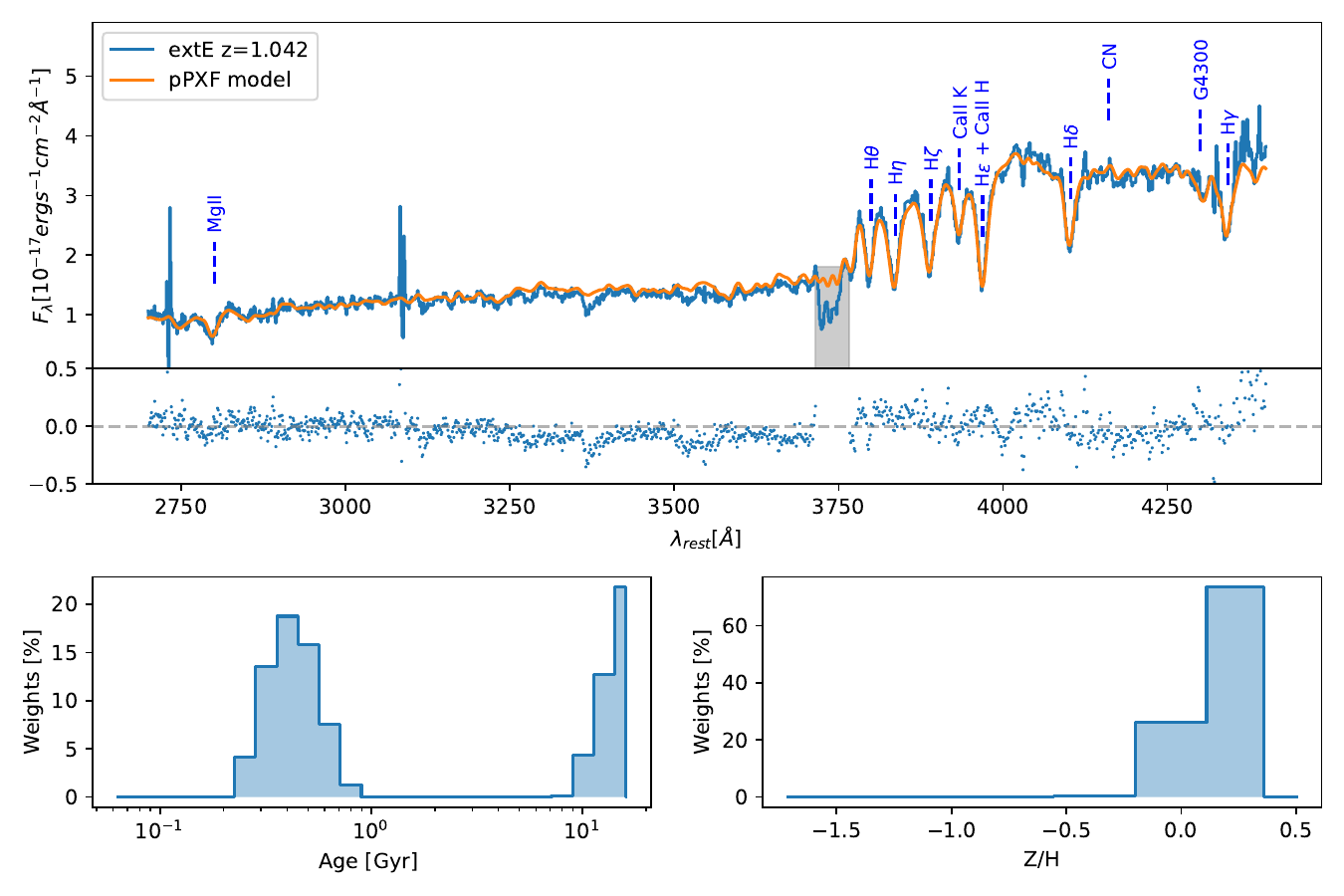}
\caption{pPXF model of the East radio source optical GTC/OSIRIS spectrum. The panel distribution is the same as in Fig.\ref{fi:ppxf_cso}. SSP templates are luminosity-weighted. The spectrum is dominated by deep absorption lines from the HII Balmer series. Given the large redshift, the UV \ion{Mg}{II} feature can be observed in the optical, its EW is indicative of recent star-formation activity.} 
\label{fi:ppxf_extE}
\end{figure*}

The spectrum was extracted at the location of the East radio source (Fig. \ref{fi:ppxf_extE}). It shows very deep H Balmer lines, and the Balmer break is also clearly visible. The \ion{Ca}{II} K \& H doublet appears asymmetric, that is, the equivalent width (EW) of the H line is about twice that of the K line. The \ion{Mg}{II} 2800\AA\, is also clearly detected, showing a measurable EW. 

The spectrum was also modelled using the pPXF software as explained in Section \ref{ssc:cso_spect}. In this case the best fit stellar population corresponds to a bimodal age distribution, a relatively young stellar population (in the range 250-700~Myr) plus an old stellar population ($> 10$~Gyr), with a solar and extrasolar metallicity (Fig. \ref{fi:ppxf_extE}). As in the case of the CSO, the optical emission can be explained with the only contribution from the host stellar population.

\subsection{The radio emission}

The high angular resolution and deep sensitivity of the radio observations allowed us to clearly separate the East radio source from the CSO at low frequencies for the first time. 
We measured the flux density of the unresolved components with the CASA task {\tt imfit}, which performs a 2D Gaussian fit in the image plane. In the case of resolved structures, we extracted the flux density from a selected polygonal region on the image plane. The error associated with the flux density is $\sigma = \sqrt{\sigma_{\rm cal}^2 + \sigma_{\rm rms}^2 + \sigma_{\rm fit}^2}$, where $\sigma_{\rm cal}$ is the uncertainty in the amplitude calibration. The term $\sigma_{\rm rms}$ is the 1-$\sigma$ noise level of the rms measured in the image plane. It depends on the ratio between the angular size of the source component, $\theta_{\rm source}$, and the beam size $\theta_{\rm beam}$, such that $\sigma_{\rm rms} = {\rm rms} \times \sqrt{\theta_{\rm source}/ \theta_{\rm beam}}$. The term $\sigma_{\rm fit}$ is the error of the Gaussian fit.
Observational parameters are reported in Tables \ref{vla-tab} and \ref{lofar-tab}. Errors associated with the spectral index are computed following the standard error propagation theory.\\

\begin{table*}
\caption{VLA observational parameters.}
\begin{center}
\begin{tabular}{cccccccc}
\hline
Component & $\nu$ & S$_{\nu}$  & S$_{\rm p, \nu}$ & $\theta_{\rm maj}$ & $\theta_{\rm min}$ & PA & Weighting\\
 & GHz & mJy & mJy/beam & arcsec & arcsec & deg  \\
 \hline
 East source & 0.368 & 21.0$\pm$2.3 & 8.3$\pm$0.7 & 6.67$\pm$0.87 & 3.17$\pm$0.49 & 86$\pm$7 & U \\
            & 0.368 & 27.7$\pm$2.0 & 11.0$\pm$0.7 & 7.7$\pm$0.5 & 3.9$\pm$0.3 & 88$\pm$3 & B \\
            & 0.352 & 30.3$\pm$4.0 & 18.3$\pm$1.8 & 9.2$\pm$1.8 & 3.2$\pm$1.3 & 90$\pm$8 & N \\
            & 1.5 & 3.2$\pm$0.4 & 3.0$\pm$0.3 & $<$1.7 & $<$1.2 & - & U \\
            & 1.5 & 6.1$\pm$0.6 & 3.3$\pm$0.3 & 4.5$\pm$0.5 & 2.5$\pm$0.6 & 83$\pm$12 & B \\
            & 1.5 & 7.6$\pm$0.6 & 4.0$\pm$0.3 & 5.7$\pm$0.5 & 3.3$\pm$0.5 & 83$\pm$9 & N \\
            & 4.8 & 1.3$\pm$0.1 & 1.2$\pm$0.1 & $<$0.14 & - & 90 & U \\
            & 4.8 & 3.6$\pm$0.3 & 1.5$\pm$0.2 & 2.3 & 1.8 & 86 & N \\
\hline
CSO & 0.368 & 25.9$\pm$1.6 & 25.1$\pm$1.4 & - & - & - & B \\
         & 1.5   & 504.0$\pm$25.0 & 504.2$\pm$25.0 & - & - & - & B \\
         & 4.8 & 1299$\pm$65 & 1299$\pm$65 & - &  - & - & U \\
\hline
\end{tabular}
\end{center}
\label{vla-tab}
\tablefoot{Column 1: Component. Column 2: Observing frequency. Columns 3 and 4: Total flux density (mJy) and peak flux density (mJy/beam). Columns 5 and 6: Deconvolved major and minor axis, respectively. Column 7: Position angle of the major axis. Column 8: Weighting used to create the image (see Table \ref{radio-oss-tab}).}
\end{table*}

\begin{table*}
\caption{LOFAR observational parameters at 150 MHz.}
\begin{center}
\begin{tabular}{cccccc}
\hline
Component & S$_{\rm 150}$  & S$_{\rm p, 150}$ & $\theta_{\rm maj}$ & $\theta_{\rm min}$ & PA \\
 & mJy & mJy/beam & arcsec & arcsec & deg  \\
 \hline
East radio source$^{a}$   & 75.4$\pm$8.0 & 34.0$\pm$3.5 & 8.7$\pm$0.4 & 4.6$\pm$0.3 & 91 \\
Core$^{b}$ & 16.7$\pm$1.7 & 13.6$\pm$1.4 & $<0.24$ & $<0.10$ & 27\\
Halo$^{b}$ & 20.5$\pm$2.3 &  -  & - & - & -   \\
\hline
CSO & 12.4$\pm$1.3 & 11.4$\pm$1.4 & $<$0.12 & $<$0.11 & - \\
\hline
\end{tabular}
\end{center}
\label{lofar-tab}
\tablefoot{Column 1: Component. Column 2: Total flux density at 150 MHz in mJy. Column 3: Peak flux density at 150 MHz in mJy/beam. Columns 4 and 5: Deconvolved major and minor axis, respectively. Column 6: Position angle of the major axis. $^a$: The integrated parameters of the East radio source are measured from LOFAR image. $^b$: The observational parameters of the core and halo components of the East radio source are measured from the ILT image.}
\end{table*}

\subsubsection{The East radio source}
 
\indent In Fig. \ref{lofar-fig} we show the images at 150 MHz of the East radio source and the CSO J0111+3906. The East radio source is clearly extended in the EW direction ($\sim 9 \times 4.6$ arcsec$^{2}$, which corresponds to $\sim$ 74$\times$38 kpc$^2$ at $z = 1.042$).
The same structure is found in the VLA P band images (Fig. \ref{vla-fig}). However, when observed with ILT, the emission of the East radio source mainly comes from the compact core region that accounts for about 22\% of the LOFAR total flux density, whereas the extended emission is almost completely resolved out. Only a hint of diffuse low-surface brightness emission is picked up by ILT observations, without any evidence of active hotspots (Fig. \ref{lofar-fig}). This structure, reminiscent of a core-halo morphology, is also seen in VLA images at 1.5 and 4.8 GHz (Figs. \ref{vla-fig} and \ref{vla-cband}). 

The synchrotron spectrum of the central region of the East radio source is a power law with $\alpha = 0.7\pm0.1$. This value was calculated using the flux density at 150 MHz from the ILT image and the VLA A-configuration image at 4.8 GHz obtained with uniform weightings and a comparable beam size. Despite the worse angular resolution of the VLA data at 1.5 GHz ($\sim 3$ arcsec), we do not see any deviation from $\alpha \sim 0.7$, indicating that there is no significant contamination from extended emission.
The spectral index between 0.15 and 4.8 GHz of the halo (without the contribution of the central region) is 0.9 $\pm$ 0.2.\\

\subsubsection{The CSO J0111+3906}
\label{result:CSO}

The CSO J0111+3906 is unresolved in all the images presented in this work, from arcsecond-scale VLA and LOFAR data to sub-arcsecond-scale ILT observations (Figs. \ref{lofar-fig}, \ref{vla-fig}, and \ref{vla-cband}). 
The spectral index between 0.37 and 1.5 GHz of the CSO is $\alpha = -2.1\pm 0.1$ in agreement with what was found by \citet{baum90} and \citet{cstan98}. Then the spectrum changes slope to $ \alpha \sim -0.8$  between 1.5 and 4.8, i.e. close to the synchrotron peak frequency, $\nu_{\rm p} \sim 4.7$ GHz \citep[e.g.][]{mo07}.

At 150 MHz, the flux density of the CSO measured in both LOFAR and ILT images is $S_{\rm 0.15} \sim 12$ mJy, indicating that all the emission arises from the central sub-arcsecond-scale compact region without any significant contribution on larger scales. This flux density is about 8.5 mJy higher than the value expected by extrapolating the flux density assuming the spectral index $\alpha_{0.37}^{1.5} \sim -2.1$ calculated between 370 and 1500 MHz. As a consequence, the spectral slope below 370 MHz changes to $\alpha_{0.15}^{0.37} = -0.9 \pm 0.1$. A similar 'flattening' of the spectral slope at low frequencies has been found in other CSOs \citep[e.g.][]{callingham17}.

\section{Discussion}
\label{discussion}

\subsection{The East radio source}
 
The detection of a galaxy behind the radio emission about 20 arcsec east of the CSO J0111+3906 poses questions about its relic nature. Deep optical photometric and spectroscopic observations proved to be fundamental in determining the nature of the East radio source.

\subsubsection{An embedded AGN}

The optical spectroscopy presented here clearly rules out a physical connection between the CSO and the East radio source. The optical object at redshift $z=1.042$ is a post-starburst galaxy with two main stellar populations: an old component (age $\leq 10$ Gyr) and a more massive younger component (0.1 Gyr $\leq$ age $\leq$ 1.0 Gyr) formed by a recent burst of star formation \citep[e.g.][]{dressler83}. The optical spectrum does not indicate the presence of an AGN, since at the redshift of the galaxy the typical AGN lines fall in the telluric absorption wavelength range ([\ion{O}{II}]) or outside the usable spectrum (e.g. [\ion{O}{III}], \ion{He}{II}), or they are not detected despite within the spectral range (e.g. [\ion{Ne}{V}]$\lambda$3426,  {H{$\beta$}).
However, since the redshift is known, the presence of an AGN may be revealed by radio data. The sub-arcsecond resolution of the ILT and VLA data pinpointed the presence of a compact core region with $\alpha \sim 0.7$ at the centre of the East radio source. We calculated the monochromatic luminosity $L_{\nu}$ by

\begin{equation}
L_{\nu} = 4 \pi D_{\rm L}^2 \frac{S_{\nu}}{(1+z)^{1 - \alpha}},
\label{lum-eq}
\end{equation}

\noindent where $D_{\rm L}$ is the luminosity distance, $S_{\nu}$ is the flux density at the frequency $\nu$, and $\alpha$ is the spectral index. The luminosity of the core region at 0.15 GHz and 1.5 GHz is $L_{\rm 0.15} = 8 \times 10^{25}$ W/Hz and $L_{\rm 1.5} = 1.5 \times 10^{25}$ W/Hz, respectively, which are much higher than those usually found in starburst galaxies \citep[e.g.][]{norris05,muxlow05}. These luminosities require an extremely high star formation rate \citep[e.g.][]{smith21}, which is inconsistent with the host stellar population, and are indicative of the presence of an AGN, despite the lack of evidence from the optical spectrum. 

Another way to infer the presence of an AGN is by determining the brightness temperature $T_{\rm b}$. We calculated the brightness temperature of the core region by

\begin{equation}
    T_{\rm b} = \left( \frac{ c^2}{2 k \nu^2} \right) \frac{S_{\nu}}{\Omega} (1 + z) \;\;,
\label{tb_eq}
\end{equation}

\noindent where $S_{\nu}$ is the flux density at frequency $\nu$, $c$ is the speed of light, $z$ is the redshift, and $k$ is the Boltzmann constant. The solid angle of the emitting region, $\Omega$, corresponds to $\Omega = \frac{\pi}{4} \theta^{2}$ where $\theta$ is the angular size. If in Eq. \ref{tb_eq} we consider the flux density at 150 MHz and $\theta < 0.1$ arcsec, we find $T_{\rm b} > 3.3 \times 10^7$ K for the central compact region. Brightness temperatures $> 10^7$ K are not readily associated with star formation, but are commonly found in AGN-related radio sources \citep[e.g.][]{lonsdale03,bondi25}. 

\subsubsection{The radio morphology}

In high-resolution VLA and ILT data, the East radio source shows a core-halo radio structure (Figs. \ref{lofar-fig}, \ref{vla-fig}, and \ref{vla-cband}). Radio sources showing core-halo morphology are rare among bright AGNs, which usually have a Fanaroff-Riley radio structure \citep[e.g.][]{fr74,laing83,morganti93,mingo19}. Although in flat-spectrum radio sources the core-halo morphology may be explained by projection effects \citep[e.g.][]{antonucci85,tadhunter16}, in steep-spectrum jetted radio sources the nature of the halo is less trivial. \\
A well-studied core-halo radio source is 3C\,317 \citep[e.g.][]{morganti93}. This source is hosted by a cD galaxy at the centre of a cool-core cluster and has a luminosity at 1.4 GHz of about $5\times 10^{24}$ W/Hz \citep{venturi04}. High-resolution very long baseline array (VLBA) observations of the central core region pointed out a flat spectrum \citep[$\alpha \sim 0$,][]{venturi00,venturi04}, suggesting that it hosts a newly born CSO, while the outer halo, with its steep spectrum \citep[$\alpha \sim 1.6$,][]{zhao93} may be the remnant of previous jet activity \citep{venturi00}. Another intriguing possibility is that the halo is caused by jet disruption due to the interaction with the cluster environment, or by the re-acceleration of old relativistic plasma by turbulence in the cooling flow \citep{venturi04}. Radio sources with core-halo morphology have been found in several cooling core clusters \citep[e.g.][]{burns90,baum91}.\\ The halo in the East radio source is not as steep as one would expect for mini-halos or radio remnants \citep[e.g.][]{gitti04,morganti17,brienza17}. New VLBA observations of the East radio source have been requested in order to characterize the pc-scale morphology of the central region and investigate its connection with the arcsecond-scale radio emission, similar to what was done for 3C\,317 \citep{venturi04}.

\subsubsection{Physical properties of the radio emission}

Apart from the central compact region resolved by ILT images, the remaining radio emission appears rather diffuse with no clear indication of compact components, such as hotspots. However, the luminosity of the entire radio source at 150 MHz is $L_{0.15} \sim 3.3\times 10^{26}$ W/Hz, slightly above the luminosity boundary between the low-power FRI and high-power FRII populations \citep{fr74}. The synchrotron spectrum of the halo (without the contribution of the central region) is well described by a straight power law with $\alpha \sim 0.9$ with no indication of significant spectral curvatures up to 4.8 GHz, which is unusual for an old plasma \citep{murgia11}. 

Although the break frequency $\nu_{\rm b}$ is undetermined by our data, we set an upper limit to the radiative age $t_{\rm rad}$ by

\begin{equation}
t_{\rm rad} = 1590 \times \left( \frac{B}{(1+z) \nu_{\rm b}} \right)^{0.5} \frac{1}{B^2 + B_{\rm CMB}^2},
\label{trad}
\end{equation}

\noindent where $t_{\rm rad}$ is in Myr and $\nu_{\rm b}$ in GHz. The magnetic field $B$, in mG, is assumed to be uniform across the source, whereas adiabatic losses are neglected \citep[e.g.][]{murgia11}. In Eq. \ref{trad}, $B_{\rm CMB}$ is the cosmic microwave background (CMB) equivalent magnetic field. It corresponds to $B_{\rm CMB} = 3.25 \times (1+z)^2$ $\mu$G, i.e. $\sim 13 \mu$G at $z=1.042$. 

We estimated the magnetic field $B$ assuming equipartition conditions and using the formula in \citet{brunetti97}

\begin{equation}
B_{\rm eq} = \left[ C{\rm ( \delta)} \frac{L_{\nu}}{V}\nu^{(\delta -1)/2} (1 + F)    \right]^{ \frac{2}{\delta + 5}} \gamma_{\rm min}^{\frac{2(2-\delta)}{\delta+5}},
\label{beq}    
\end{equation}

\noindent where $\delta$ is defined as $\delta = (\alpha + 1)/2$ and is related to the electron energy distribution such as $N(\gamma) \propto \gamma^{- \delta}$, whereas F is the ratio between the energy densities of relativistic protons and electrons, which is $F=0$ for a pure electron-positron plasma. Representative values of the function $C(\delta)$ are tabulated in \citet{brunetti97}. In our computation of $B_{\rm eq}$ we assumed $\delta = 2.9$, which corresponds to $\alpha=0.95$, i.e. the spectral index of the integrated synchrotron spectrum of the East radio source. We approximate the volume $V$ of the radio source to a prolate ellipsoid that is homogeneously filled by relativistic plasma

\begin{equation}
V = \frac{\pi}{6} \left( \frac{D_{L}}{(1+z)^2}\right)^3 \theta_{\rm maj} \theta_{\rm min}^2,
\label{volume}
\end{equation}

\noindent where $\theta_{\rm maj}$ and $\theta_{\rm min}$ are the major and minor angular sizes, respectively, and $D_{L}$ is the luminosity distance at the redshift $z$.
If in Eq. \ref{beq} we consider $\gamma_{\rm min} = 100$, the luminosity at 150 MHz, and the volume obtained from Eq. \ref{volume} assuming the angular size derived from the LOFAR image, we obtain $B_{\rm eq} \sim 20 \; \mu G$. 

We can now set an upper limit to the radiative age of the East radio source by inserting $B_{\rm eq}$ and $B_{\rm CMB}$ into Eq. \ref{trad}. Since the spectrum does not show any steepening up to the highest frequency sampled by our observations, we conservatively assume $\nu_{\rm b} = 4.8$ GHz. We end up with $t_{\rm rad} <4\times 10^6$ yr, which is typical of medium-sized compact steep-spectrum (CSS) young radio galaxies \citep[e.g.][]{fanti95,murgia99}. If the onset of radio emission and the starburst episode are causally connected, there must be a long time gap between the two events ($\sim 10^8$ yr). Significant time-delay between starburst and radio-AGN activity has been found in other young radio sources, suggesting multiple galaxy encounters (t $\sim 10^{5 - 6}$ yr) as the trigger of jet production, rather than a merger event \citep[t $\sim 10^9$ yr; e.g.][]{emonts06}. The object detected about 3" south-east of the host galaxy might be responsible for both the starburst and radio emission, but additional optical data are needed to prove this scenario.

\subsection{The CSO J0111+3906}

The optical spectrum of the CSO is typical of a passive evolving galaxy whose optical light is dominated by an old stellar population that originated in a single burst, whereas the AGN contribution is negligible. These characteristics have been found in other young radio galaxies \citep[e.g.][]{snellen99,labiano07}. 
Although our optical observations exclude a connection between the CSO and the East radio source, the change in the spectral slope observed at 150 MHz keeps the door open to the recurrence scenario.  
If in Eq. \ref{lum-eq} we assume the flux density excess of 8.5 mJy found at 150 MHz (see Sect. \ref{result:CSO}), we obtain $L_{150} \sim 1.5 \times 10^{25}$ W/Hz, which cannot be reconciled with stellar emission and might be due to old plasma from a past activity \citep[e.g.][]{callingham17}. 
The consistency (within the errors) between the flux density measured by LOFAR and that measured in ILT images indicates that this fossil emission is confined within the host galaxy ($\theta < 0.1$ arcsec, i.e. $<$710 pc at the redshift of the source; see Table \ref{lofar-tab}). If we consider this excess as a remnant, its compact linear size suggests that the previous epoch of radio emission switched off not long after its trigger. If we assume a constant advance hotspot speed of (0.1 - 0.2)$c$ \citep[e.g.][]{polatidis03,an12b}, we obtain a kinematic age between 6$\times$10$^{3}$ and 1.2$\times$10$^{4}$ yr. Owing to the large uncertainties in the source parameters and the many assumptions made, we note that the exact value is not relevant, but rather the order of magnitude, which suggests a relatively short duty-cycle of about 10$^{4}$ yr for J0111+3906. An alternative interpretation is that radio excess traces aged electrons that fill the volume previously swept by either a disrupted jet, or a precessing or laterally drifting jet, as expected in the jet redirection scenario \citep{cstan25}. 
However, to confirm the presence of relic plasma associated with the CSO we need high-resolution ($\leq 6$ arcsec) observations below 150 MHz to determine the spectral slope at very low frequencies and set a stringent upper limit to the source size.

\section{Conclusions}
\label{summary}

We presented results on GTC/OSIRIS+ optical spectroscopy and radio VLA and LOFAR observations of the radio emission about 20 arcseconds east of the CSO radio galaxy J0111+3906. The conclusions we can draw from this study are as follows: 

\begin{itemize}

    \item The redshift of the CSO radio galaxy J0111+3906 is confirmed to be $z=0.668$. The optical spectrum of the CSO represents an old stellar population and superimposed a few emission lines that correspond to low ionization gas emission.
    
    \item The optical spectrum of the East radio source corresponds to a galaxy at redshift $z=1.042$. This fact disproves any physical relationship between the two targets. The spectrum can be explained as produced by a stellar population with a bimodal age distribution, a young population with mean stellar age of 400 Myr, and an old one with mean stellar age of 14 Gyr.
     
    \item The high angular resolution of VLA A-configuration and ILT observations allows us to resolve the radio emission of the East radio source in a compact region, centred on the host galaxy, enshrouded by a diffuse halo. Despite the lack of typical AGN lines in the optical spectrum, the presence of an AGN is revealed by radio luminosity exceeding 10$^{25}$ W/Hz and brightness temperatures $> 10^{7}$ K, which are not readily associated with star formation.
    
    \item The radio spectrum of the core region of the East radio source is 0.7$\pm$0.1 between 150 MHz and 4.8 GHz, while that of the halo is $\sim$0.9 with no evidence of spectral curvature. Assuming equipartition conditions, we derived a magnetic field of 20 $\mu$G, and set an upper limit to the radiative age of 4$\times$10$^6$ yr, which makes the East radio source part of the sub-population of CSS young radio sources. If the onset of radio emission and the burst of star formation are causally connected, there must be a long time gap between the two phenomena, which may be explained by multiple galactic interactions rather than a major merger event.
    
\item Although our study disproves that the kpc-scale emission is a remnant of a past activity of the CSO radio galaxy J0111+3906, the flux density excess at 150 MHz keeps the door open to the presence of fossil plasma on scales of hundreds of parsecs. 

\end{itemize}

Despite decades of investigation, the duty-cycle of radio emission is still a key issue in understanding the formation of jetted AGNs. 
The improvement in sensitivity and resolution at low frequencies is providing a step forward in our understanding of the life-cycle of radio galaxies. However, the same studies on compact and young radio sources prove to be challenging.
The improved sensitivity and (sub-)arcsecond resolution at very low frequencies ($<$360 MHz) will be crucial in disclosing fossil plasma from past activity epochs. Given their inverted spectrum below a few GHz, CSOs are the best candidates to search for remnant emission from past activity, by looking at the presence of flux density excess at very low frequencies. However, flux density excess may be due to blending of sources, and optical/infrared spectroscopic observations are thus crucial to establish the nature of the radio emission.

\begin{acknowledgements}

We thank the anonymous referee for reading the manuscript carefully and making valuable suggestions.
The VLA is operated by the US National Radio Astronomy Observatory which is a facility of the National Science Foundation operated under cooperative agreement by Associated Universities, Inc. This work has made use of the NASA/IPAC Extragalactic Database (NED) which is operated by the JPL, California Institute of Technology, under contract with the National Aeronautics and Space Administration. 
MO and FD acknowledge financial support from INAF 2022 and 2024 fundamental research programmes ob. fun. 1.05.12.05.15 and 1.05.24.02.09. EDR acknowledges support by the Deutsche Forschungsgemeinschaft (DFG) and by the Fondazione ICSC, Spoke 3 Astrophysics and Cosmos Observations. National Recovery and Resilience Plan (Piano Nazionale di Ripresa e Resilienza, PNRR) Project ID CN\_00000013 \enquote{Italian Research Center for High-Performance Computing, Big Data and Quantum Computing} funded by MUR Missione 4 Componente 2 Investimento 1.4: Potenziamento strutture di ricerca e creazione di \enquote{campioni nazionali di R\&S (M4C2-19)} - Next Generation EU (NGEU). 
Based on observations made with the GTC telescope, in the Spanish Observatorio del Roque de los Muchachos of the Instituto de Astrofísica de Canarias, under Director’s Discretionary Time.
The reference of GTC/DDT proposal is GTC2024-253(GTC01-24BDDT according to GTC).
This manuscript is based on data obtained with the International LOFAR Telescope (ILT). LOFAR~\citep{vanhaarlem2013} is the Low Frequency Array designed and constructed by ASTRON. It has observing, data processing, and data storage facilities in several countries, which are owned by various parties (each with their own funding sources), and which are collectively operated by the ILT foundation under a joint scientific policy. The ILT resources have benefited from the following recent major funding sources: CNRS-INSU, Observatoire de Paris and Université d’Orléans, France; BMBF, MIWF-NRW, MPG, Germany; Science Foundation Ireland (SFI), Department of Business, Enterprise and Innovation (DBEI), Ireland; NWO, The Netherlands; The Science and Technology Facilities Council, UK; Ministry of Science and Higher Education, Poland; The Istituto Nazionale di Astrofisica (INAF), Italy. This research made use of the LOFAR-IT computing infrastructure supported and operated by INAF, including the resources within the PLEIADI special \enquote{LOFAR} project by USC-C of INAF, and by the Physics Dept. of Turin University (under the agreement with Consorzio Interuniversitario per la Fisica Spaziale) at the C3S Supercomputing Centre, Italy.

\end{acknowledgements}

\begin{appendix}

\onecolumn
\section{VLA images}

\FloatBarrier

\begin{table}
\caption{Log of VLA observations.}
\begin{center}
\begin{tabular}{lcccc}
\hline
Code & Band & Array & Obs. time & Date \\
\hline
&&&&\\
VLA/24B-144 & P & A & 75 &10 Dec 2024\\
VLA/24B-144& P & A & 75 &29 Dec 2024\\
VLA/24B-144& P & A & 75 & 4 Jan 2025\\
VLA/23A-006 & L & A &15 &15 Jan 2023\\
V085A       & C & A & 540 & 06 Aug 1999\\
GG15        & C & B & 38 & 25 Feb 1993\\
AC600       & C & C & 18 & 13 Jul 2001\\
AP380       & C & D & 3 & 6 Aug 2000\\
AY102       & C & D & 7 & 28 Mar 1999\\
\hline
\end{tabular}
\end{center}
\tablefoot{Column 1: Project code. Column 2: Observing band. Column 3: Array configuration. Column 4: On-source time in minutes. Column 5: Observation date.}
\label{vla-log}
\end{table}

\FloatBarrier

\twocolumn

\begin{figure*}
\begin{center}
\includegraphics[width=0.6\columnwidth]{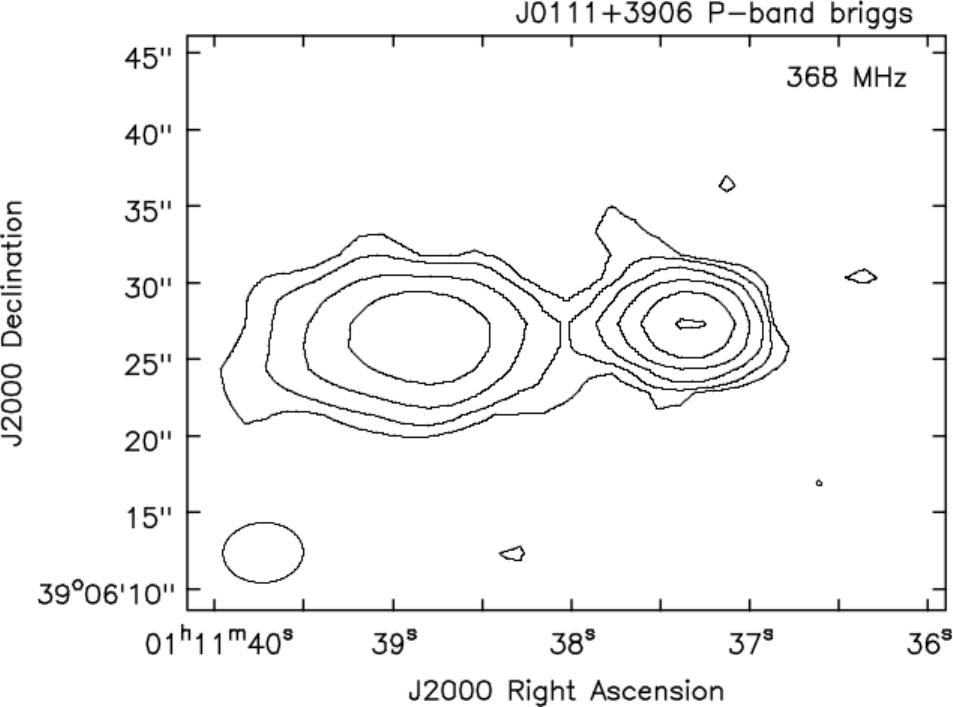}
\includegraphics[width=0.6\columnwidth]{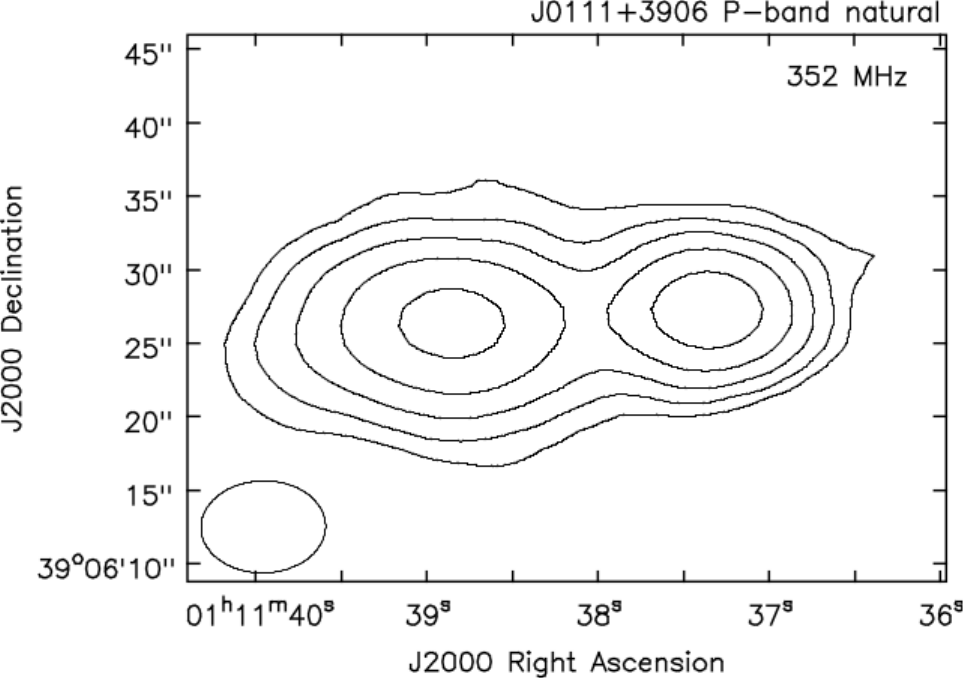}
\includegraphics[width=0.6\columnwidth]{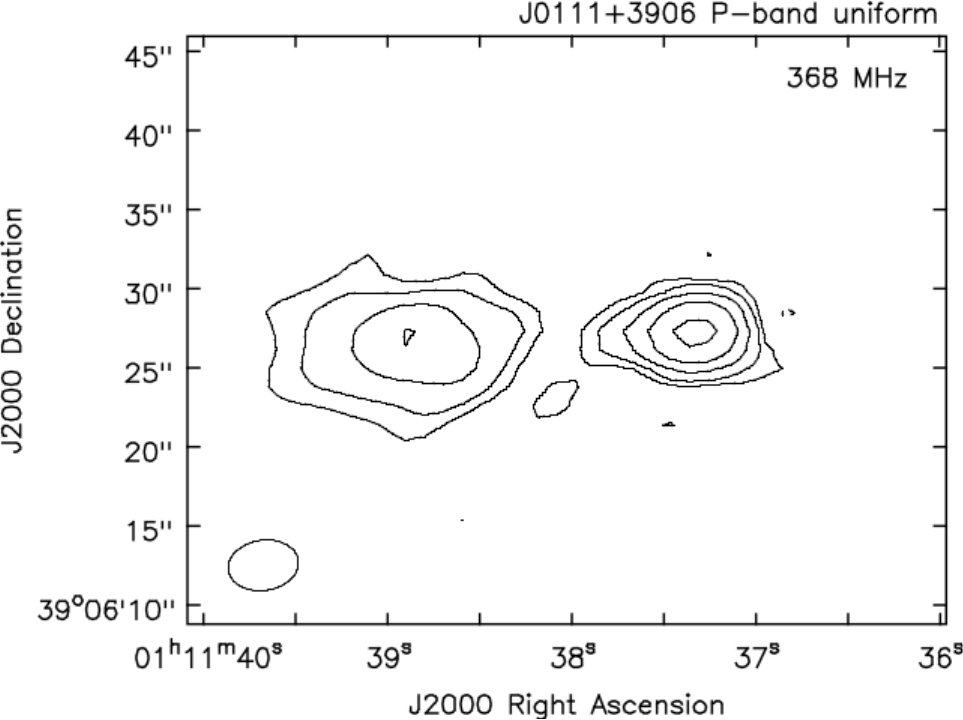}
\includegraphics[width=0.6\columnwidth]{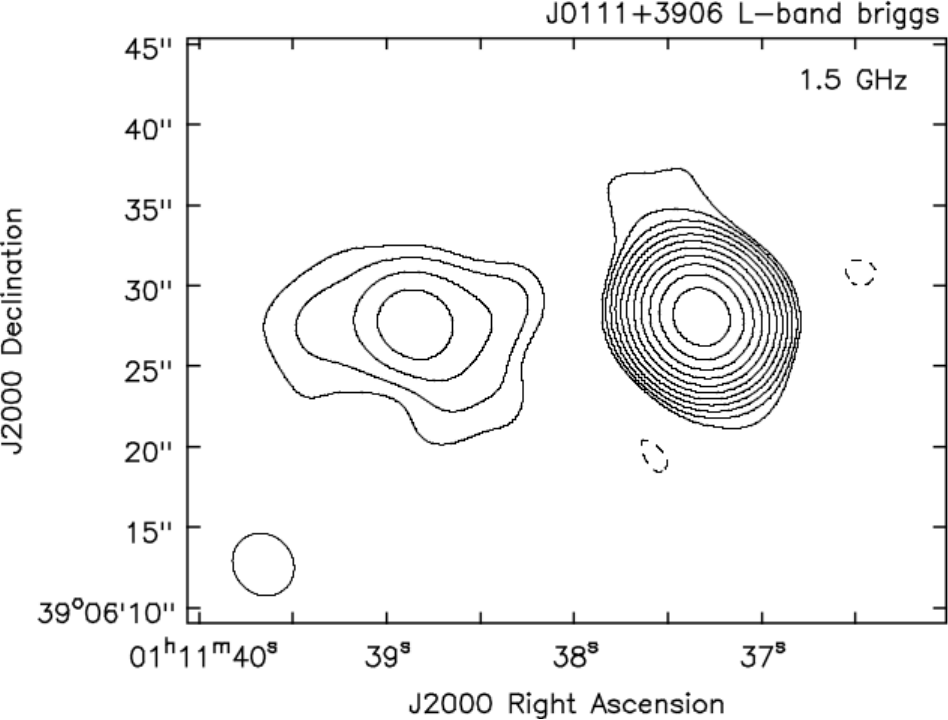}
\includegraphics[width=0.6\columnwidth]{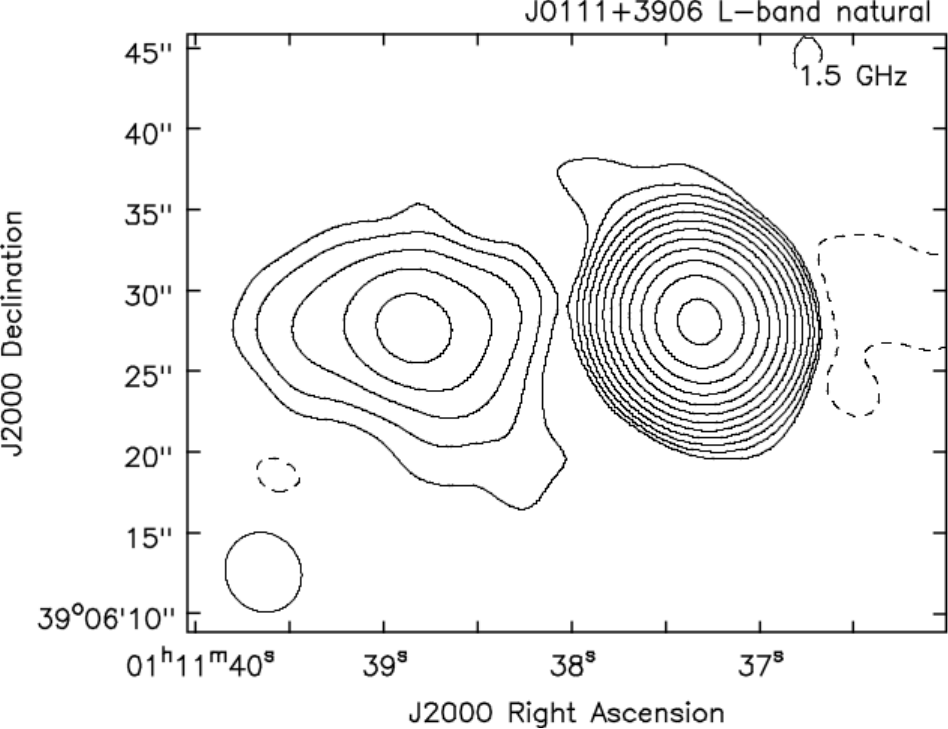}
\includegraphics[width=0.6\columnwidth]{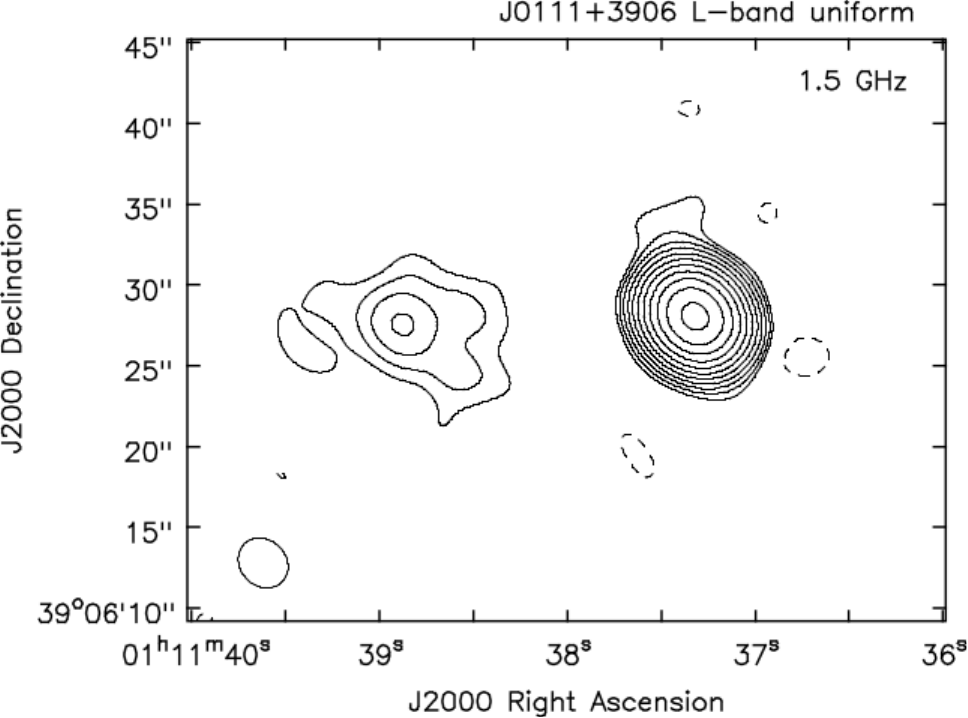}
\caption{VLA images in P band (upper panel) and L band (lower panel). On each image, we provide the observing frequency and weightings. Image parameters are reported in Table \ref{radio-oss-tab}. The first contour is three times the rms. Contours increase by a factor of 2. The restoring beam is plotted in the bottom left-hand corner of each image.}
\label{vla-fig}
\end{center}
\end{figure*}

\begin{figure*}
\begin{center}
\includegraphics[width=0.6\columnwidth]{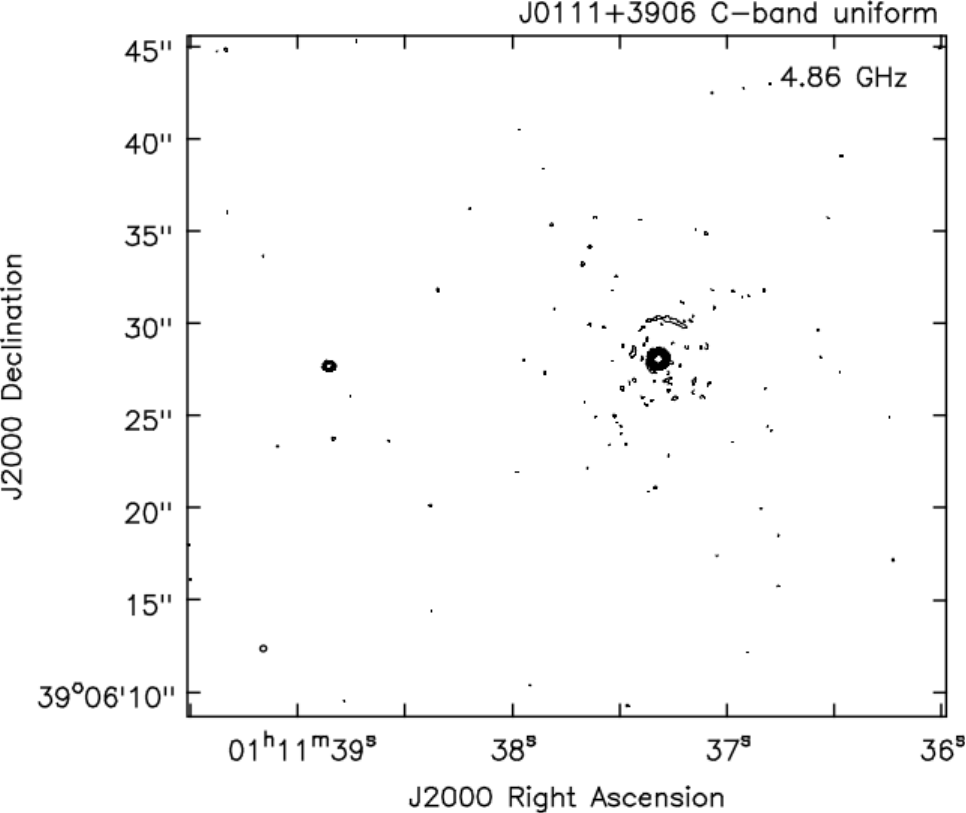}
\includegraphics[width=0.6\columnwidth]{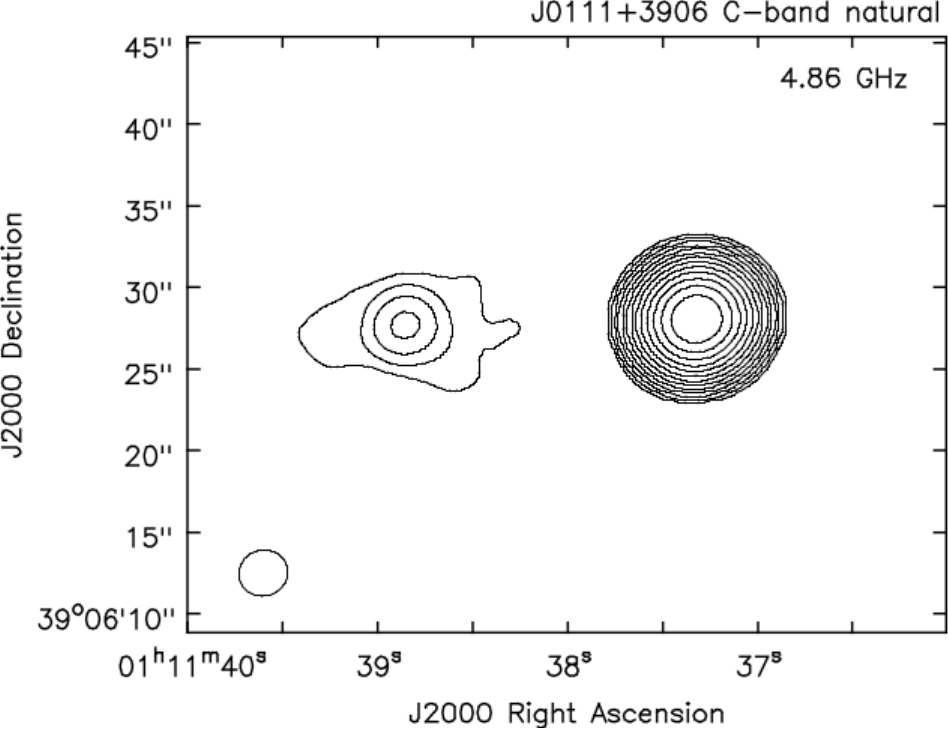}
\caption{Images at 4.86 GHz of J0111+3906 taken with the historical VLA in A-configuration (left-hand panel) and combining all the arrays (right-hand panel). Image parameters are reported in Table \ref{radio-oss-tab}. The first contour is three times the rms. Contours increase by a factor of 2. The restoring beam is plotted in the bottom left-hand corner of each image.}
\label{vla-cband}
\end{center}
\end{figure*}

\end{appendix}


\begin{thebibliography}{}

\bibitem[An \& Baan(2012a)]{an12a}
An, T., Baan, W.A. 2012a, ApJ, 760, 77

\bibitem[An et al.(2012b)]{an12b}
An, T., Wu, F., Yang, J., et al. 2012b, ApJS, 198, 5

\bibitem[Antonucci \& Ulvestad(1985)]{antonucci85}
Antonucci, R.R.J., Ulvestad, J.S. 1985, ApJ, 294, 158

\bibitem[Baum et al.(1990)]{baum90}
Baum, S.A., O'Dea, C.P., Murphy, D.W., de Bruyn, A.G. 1990, A\&A, 232, 19

\bibitem[Baum \& O'Dea(1991)]{baum91}
Baum, S.A., O'Dea, C.P. 1991, MNRAS, 250, 737

\bibitem[Bondi et al.(2025)]{bondi25}
Bondi, M., Prandoni, I., Magliocchetti, M., et al. 2025, A\&A, 698, 275

\bibitem[Brienza et al.(2017)]{brienza17}
Brienza, M., Godfrey, L., Morganti, R., et al. 2017, A\&A, 606, 98

\bibitem[Brocksopp et al.(2007)]{brocksopp07}
Brocksopp, C., Kaiser, C.R., Schoenmakers, A.P., de Bruyn, A.G. 2007, MNRAS, 382, 1019

\bibitem[Brunetti et al.(1997)]{brunetti97}
Brunetti, G., Setti, G., Comastri, A. 1997, A\&A, 325, 898

\bibitem[Burns(1990)]{burns90}
Burns, J.O. 1990, AJ, 99, 14

\bibitem[Callingham et al.(2017)]{callingham17}
Callingham, J.R., Ekers, R.D., Gaensler, B.M., et al. 2017, ApJ, 836, 174

\bibitem[\protect\citeauthoryear{Cappellari}{2023}]{Cappellari2023} 
Cappellari M., 2023, MNRAS, 526, 3273. doi:10.1093/mnras/stad2597

\bibitem[Cepa et al.(2003)]{Cepa03}
Cepa, J., Aguiar-Gonzalez, M., Bland-Hawthorn, J., et al. 2003, in SPIE Conference Series, Vol. 4841, Instrument Design and Performance for Optical/Infrared Ground-based Telescopes, ed. M. Iye \& A. F. M. Moorwood, 1739–1749

\bibitem[Czerny et al.(2009)]{czerny09}
Czerny, B., Siemiginowska A., Janiuk A., Nikiel-Wroczynski B., Stawarz, \L, 2009, ApJ, 698, 840

\bibitem[de Gasperin et al.(2019)]{degasperin2019}
de Gasperin, F., Dijkema, T. J., Drabent, A., et al. 2019, A\&A, 622, A5

\bibitem[Dressler \& Gunn(1983)]{dressler83}
Dressler, A., Gunn, J.E. 1983, ApJ, 270, 7

\bibitem[Emonts et al.(2006)]{emonts06}
Emonts, B.H.C., Morganti, R., Tadhunter, C.N., et al. 2006, A\&A, 454, 125

\bibitem[Fanaroff \& Riley(1974)]{fr74}
Fanaroff, B.L., Riley, J.M. 1974, MNRAS, 167, 31

\bibitem[Fanti et al.(1995)]{fanti95}
Fanti, C., Fanti, R., Dallacasa, D., et al. 1995, A\&A, 302, 31

\bibitem[Ficarra et al.(1985)]{ficarra1985}
Ficarra, A., Grueff, G., Tomassetti, G. 1985, A\&AS, 59, 255

\bibitem[\protect\citeauthoryear{Girardi et al.}{2000}]{Girardi2000} Girardi L., Bressan A., Bertelli G., Chiosi C., 2000, A\&AS, 141, 371. doi:10.1051/aas:2000126

\bibitem[Gitti et al.(2004)]{gitti04}
Gitti, M., Brunetti, G., Feretti, L., Setti, G. 2004, A\&A, 417, 1

\bibitem[Gordon et al.(2023)]{Gordon23} 
Gordon, K.~D., Clayton, G.~C., Decleir, M., et al.\ 2023, \apj, 950, 2, 86.

\bibitem[Gugliucci et al.(2005)]{gugliucci05}
Gugliucci, N.E., Taylor, G.B., Peck, A.B., Giroletti, M. 2005, ApJ, 622, 136

\bibitem[Intema et al.(2017)]{intema2017}
Intema, H. T., Jagannathan, P., Mooley, K. P., Frail, D. A. 2017, A\&A, 598, A78

\bibitem[Jackson et al.(2016)]{jackson2016}
Jackson, N., Tagore, A., Deller, A., et al. 2016, A\&A, 595, A86

\bibitem[Jackson et al.(2022)]{jackson2022}
Jackson, N., Badole, S., Morgan, J., et al. 2022, A\&A, 658, A2

\bibitem[Kennicutt(1998)]{Kennicut98} 
Kennicutt R.~C., 1998, ApJ, 498, 541.

\bibitem[Kiehlmann et al.(2024)]{kiehlmann24}
Kiehlmann, S., Readhead, A.C.S., O'Neil, S., et al. 2024, ApJ, 961, 241

\bibitem[Ku\'zmicz et al.(2017)]{kuzmicz17}
Ku\'zmicz, A., Jamrozy, M., Kozie{\l}-Wierzbowska, D., We{\.z}gowiec, M. 2017, MNRAS, 471, 3805

\bibitem[Labiano et al.(2007)]{labiano07}
Labiano, A., Barthel, P.D., O'Dea, C.P., et al. 2007, A\&A, 463, 97

\bibitem[Lacy et al(2020)]{lacy2020}
Lacy, M., Baum, S. A., Chandler, C. J., et al. 2020, PASA, 132, 035001

\bibitem[Laing et al.(1983)]{laing83}
Laing, R.A., Riley, J.M., Longair, M.S. 1983, MNRAS, 204, 151

\bibitem[Lara et al.(1999)]{lara99}
Lara, L., M\'arquez, I., Cotton, W.D., et al. 1999, A\&A, 348, 699

\bibitem[Lawrence et al.(1996)]{lawrence96}
Lawrence, C.R., Zucker, J.R., Readhead, A.C.S., et al. 1996, ApJS, 107, 541

\bibitem[Lonsdale et al.(2003)]{lonsdale03}
Lonsdale, C.J., Lonsdale, C.J., Smith, H.E., Diamond, P.J. 2003, ApJ, 592, 804

\bibitem[McMullin et al.(2007)]{mcmullin07}
McMullin J.P., Waters B., Schiebel D., Young W., Golap K. 2007, in Shaw
R.A., Hill F., Bell D.S., eds, ASP Conf. Ser. Vol. 376, Astromical Data
Analysis and Systems XVI. Astron. Soc. Pac., San Francisco, p. 127

\bibitem[Mingo et al.(2019)]{mingo19}
Mingo, B., Croston, J.H., Hardcastle, M.J., et al. 2019, MNRAS, 488, 2701

\bibitem[Morabito et al.(2022)]{morabito2022}
Morabito, L. K., Jackson, N. J., Mooney, S., et al. 2022, A\&A, 658, A1

\bibitem[Morganti et al.(1993)]{morganti93}
Morganti, R., Killeen, N.E.B., Tadhunter, C.N. 1993, MNRAS, 263, 1023

\bibitem[Morganti(2017)]{morganti17}
Morganti R., 2017, Nat. Astron., 1, 596

\bibitem[Murgia et al.(1999)]{murgia99}
Murgia, M., Fanti, C., Fanti, R., et al. 1999, A\&A, 345, 769

\bibitem[Murgia(2003)]{murgia03}
Murgia, M. 2003, PASA, 20,19

\bibitem[Murgia et al.(2011)]{murgia11}
Murgia, M., Parma, P., Mack, K.-H., et al. 2011, A\&A, 526, 148

\bibitem[Muxlow et al.(2005)]{muxlow05}
Muxlow, T.W.B., Richards, A.M.S., Garrington, S.T., et al. 2005, MNRAS, 358, 1159

\bibitem[Norris et al.(2005)]{norris05}
Norris, R.P., Huynh, M.T., Jackson, C.A., et al. 2005, AJ, 130, 1358

\bibitem[O'Dea \& Saikia(2021)]{odea21}
O'Dea, C.P., Saikia, D.J. 2021, A\&AR, 29, 1

\bibitem[Offringa et al.(2014)]{offringa2014}
Offringa, A. R., McKinley, B., Hurley-Walker, N., et al. 2014, MNRAS, 444, 606

\bibitem[Offringa \& Smirnov(2017)]{offringa2017}
Offringa, A. R., Smirnov, O. 2017, MNRAS, 471, 301

\bibitem[Orienti et al.(2007)]{mo07}
Orienti, M., Dallacasa, D., Stanghellini, C. 2007, A\&A, 475, 813

\bibitem[Orienti \& Dallacasa(2008)]{mo08}
Orienti, M., Dallacasa, D. 2008, A\&A, 487, 885

\bibitem[Orienti et al.(2010)]{mo10}
Orienti, M., Murgia, M., Dallacasa, D. 2010, MNRAS, 402, 1892

\bibitem[Orienti et al.(2023)]{mo23}
Orienti, M., Murgia, M., Dallacasa, D., Migliori, G., D'Ammando, F. 2023, MNRAS, 522, 3877

\bibitem[Orr\'u et al.(2010)]{orru10}
Orr\'u, E., Murgia, M., Feretti, L., et al. 2010, A\&A, 515, 50

\bibitem[Owsianik et al.(1998) ]{owsianik98}
Owsianik, I., Conway, J.E., Polatidis, A.G. 1998, A\&A, 336L, 37

\bibitem[Pennell, Runnoe, \& Brotherton(2017)]{Pennell17} 
Pennell A., Runnoe J.~C., Brotherton M.~S., 2017, MNRAS, 468, 1433

\bibitem[Perley \& Butler(2017)]{pb17}
Perley, R.A., Butler, B.J. 2017, ApJS, 230, 7

\bibitem[Polatidis \& Conway(2003)]{polatidis03}
Polatidis, A.G., Conway, J.E. 2003, PASA, 20,69

\bibitem[Prochaska et al.(2020a)]{Prochaska20a}
Prochaska, J.X., Hennawi, J.F., Kyle,  B.W., et al. 2020, JOSS, 5, 2308 

\bibitem[Prochaska et al.(2020b)]{Prochaska20b}
Prochaska, J.X., Hennawi, J.F., Cooke,  R., et al. 2020, Zenodo, 3743493

\bibitem[Readhead et al.(1996)]{readhead96}
Readhead, A.C.S., Taylor, G.B., Pearson, T.J., Wilkinson, P.N. 1996, ApJ, 460, 634

\bibitem[Readhead et al.(2024)]{readhead24}
Readhead, A.C.S., Ravi, V., Blandford, R.D., et al. 2024, ApJ, 961, 242

\bibitem[Scaife \& Heald(2012)]{scaife12}
Scaife, A.M.M., Heald, G.L. 2012, MNRAS, 423, L30

\bibitem[Shimwell et al.(2022)]{shimwell2022}
Shimwell, T. W., Hardcastle, M. J., Tasse, C., et al. 2022, A\&A, 659, A1

\bibitem[Schoenmakers et al.(2000)]{schoenmakers00}
Schoenmakers, A.P., de Bruyn, A.G., R\"ottgering, H.J.A., van der Laan, H., Kaiser, C.R. 2000, MNRAS, 315, 371

\bibitem[Smith et al.(2021)]{smith21}
Smith, D.J.B., Haskell, P., G\"urkan, G., et al. 2021, A\&A, 648, 6

\bibitem[Snellen et al.(1999)]{snellen99}
Snellen, I.A.G., Schilizzi, R.T., Bremer, M.N., et al. 1999, MNRAS, 307, 149

\bibitem[Stanghellini et al.(1993)]{cstan93}
Stanghellini, C., O'Dea, C.P., Baum, S.A., Laurikainen, E. 1993 ApJS, 88, 1

\bibitem[Stanghellini et al.(1998)]{cstan98}
Stanghellini, C., O'Dea, C.P., Dallacasa, D., et al. A\&AS, 131, 303

\bibitem[Stanghellini et al.(2005)]{cstan05}
Stanghellini, C., O'Dea, C.P., Dallacasa, D., et al. 2005, A\&A, 443, 891

\bibitem[Stanghellini et al.(2025)]{cstan25}
Stanghellini, C., Orienti, M., Spingola, et al. 2025, A\&A, 695, A179

\bibitem[Tadhunter(2016)]{tadhunter16}
Tadhunter, C. 2016, A\&ARv, 24, 10

\bibitem[van Haarlem et al.(2013)]{vanhaarlem2013}
van Haarlem, M. P., Wise, M. W., Gunst, A. W., et al. 2013, A\&A, 556, A2

\bibitem[van Weeren et al.(2016)]{vanweeren2016}
van Weeren, R. J., Brunetti, G., Br\"uggen, M., et al. 2016, ApJ, 818, 204

\bibitem[van Weeren et al.(2021)]{vanweeren2021}
van Weeren, R. J., Shimwell, T. W., Botteon, A., et al. 2021, A\&A, 651, A115

\bibitem[\protect\citeauthoryear{Vazdekis et al.}{2016}]{Vazdekis2016} Vazdekis A., Koleva M., Ricciardelli E., R{\"o}ck B., Falc{\'o}n-Barroso J., 2016, MNRAS, 463, 3409. doi:10.1093/mnras/stw2231

\bibitem[Venturi et al.(2000)]{venturi00}
Venturi, T., Morganti, R., Tzioumis, T., Reynolds, J. 2000, A\&A, 363, 84

\bibitem[Venturi et al.(2004)]{venturi04}
Venturi, T., Dallacasa, D., Stefanachi, F. 2004, A\&A, 422, 515

\bibitem[Williams et al.(2016)]{williams2016}
Williams, W. L., van Weeren, R. J., R\"ottgering, H. J. A., et al. 2016, MNRAS, 460, 2385

\bibitem[Zhao et al.(1993)]{zhao93}
Zhao, J.-H., Sumi, D. M., Burns, J.O., Duric, N., 1993, ApJ, 416, 51

\end{thebibliography}
\end{document}